\documentclass[11pt]{article}
\usepackage[centertags]{amsmath}
\usepackage{amssymb}
\usepackage{graphicx}
\usepackage{cite}

\newcommand\mycite[1]{\raisebox{0.3em}{\tiny\cite{#1}}}

\renewcommand{\baselinestretch}{1.4}

\begin{document}
\title{\bf Sensitivity Amplification in the Phosphorylation-Dephosphorylation Cycle:\\\Large{Nonequilibrium steady states, chemical master equation and temporal
cooperativity}}

\author{Hao Ge\footnote{LMAM, School of Mathematical Sciences, Peking University, Beijing 100871,
P.R.China; email: edmund\_ge@tom.com} \and Min Qian\footnote{LMAM,
School of Mathematical Sciences, Peking University, Beijing 100871,
P.R.China;}}

\maketitle
\date{}

\begin{abstract}
A new type of cooperativity termed temporal cooperativity [Biophys.
Chem. 105 585-593 (2003), Annu. Rev. Phys. Chem. 58 113-142 (2007)],
emerges in the signal transduction module of
phosphorylation-dephosphorylation cycle (PdPC). It utilizes multiple
kinetic cycles in time, in contrast to allosteric cooperativity that
utilizes multiple subunits in a protein. In the present paper, we
thoroughly investigate both the deterministic (microscopic) and
stochastic (mesoscopic) models, and focus on the identification of
the source of temporal cooperativity via comparing with allosteric
cooperativity.

A thermodynamic analysis confirms again the claim that the chemical
equilibrium state exists if and only if the phosphorylation
potential $\triangle G=0$, in which case the amplification of
sensitivity is completely abolished. Then we provide comprehensive
theoretical and numerical analysis with the first-order and
zero-order assumptions in phosphorylation-dephosphorylation cycle
respectively. Furthermore, it is interestingly found that the
underlying mathematics of temporal cooperativity and allosteric
cooperativity are equivalent, and both of them can be expressed by
``dissociation constants'', which also characterizes the essential
differences between the simple and ultrasensitive PdPC switches.
Nevertheless, the degree of allosteric cooperativity is restricted
by the total number of sites in a single enzyme molecule which can
not be freely regulated, while temporal cooperativity is only
restricted by the total number of molecules of the target protein
which can be regulated in a wide range and gives rise to the
ultrasensitivity phenomenon.
\begin{flushleft}
\end{flushleft}
\end{abstract}


\section{Introduction}

Biological signal transduction processes are increasingly understood
in quantitative terms, such that the switching of enzymes and
proteins between phosphorylated and dephosphorylated states becomes
a universal module \mycite{Fisch71,Kreb71}. The biological activity
of a target protein is often wakened by the phosphorylation reaction
catalyzed by a specific kinase, and restrained by the
dephosphorylation reaction catalyzed by a specific phosphatase,
which is quite similar to the turning on and off procedure of an
ordinary switch.

One of the key concepts in PdPC signaling is the switching
sensitivity: the sharpness of the activation of the substrate
protein in response to the concentration of the kinase is basic in
the perspective of metabolic control analysis, usually termed as
Hill coefficient first proposed by A. V. Hill \mycite{HillAV}.

Actually, the research about the sensitivity of single-enzyme
catalysis activity, also known as the allosteric cooperativity, has
already been developed for about forty years, since the classic
paper of Monod, Wyman and Changeux  \mycite{MWC} and Koshland,
Nemethy and Filmer \mycite{KNF}. It is found in experiments that
very few individual enzymes show positive cooperativity with Hill
coefficient greater than $4$.

However, in the case of multi-enzyme systems such as the
phosphorylation-dephosphorylation module, the situation is quite
different. In the early 1980s, Goldbeter and Koshland
 \mycite{GK,KGS} discovered the ultrasensitivity phenomenon of a PdPC switch
in terms of the zeroth order kinetics of kinase and phosphatase,
where the Hill coefficient can be extremely high. Moreover, it has
already been observed in experiments \mycite{HF}.

Recently, Qian \mycite{QH2003,QH2007} has further elucidated the
importance of open-system chemical reaction in terms of continuous
ATP hydrolysis. It was found that the thermodynamic energy aspect of
the signal transduction plays an important role in further
understanding the function of PdPC switches, which confirms the
well-known belief that signal transduction in biological systems
actually consumes energy \mycite{NP}.

Most of the previous models \mycite{SC,GK,QH2003,HNR} built for the
phosphorylation and dephosphorylation module were traditionally
based on deterministic, coupled nonlinear ordinary differential
equations in terms of regulatory mechanisms and kinetic parameters,
which are widely used in the field of computational biology
\mycite{Mu,FMWT}. Nowadays, as there is a growing awareness of the
basic character of noise in the study of the effects of noise in
biological networks, it becomes more and more important to develop
stochastic models with chemical master equations (CME) based on
biochemical reaction stoichiometry, molecular numbers, and kinetic
rate constants. Such an approach has already provided important
insights and quantitative characterizations of a wide range of
biochemical systems \mycite{Gi76,Mc1963,Mc1964,Mc,GHO,Van,QH1,Zhou},
especially in recent studies on gene expression \mycite{SES,ZMTW}.

On the other hand, these stochastic models for systems cell biology
would exhibit nonequilibrium steady states (NESS), in which their
mesoscopic properties can be rigorously investigated from the
trajectory point of view \mycite{hqjpcm05,hqjpc06,GQQ_MBS2007}.
Moreover, several recently interesting experimental results can only
be explained by stochastic models \mycite{Ge_JPC2007}.

The aim of this paper is to thoroughly investigate temporal
cooperativity \mycite{QH2003} emerged in the signal transduction
module of phosphorylation-dephosphorylation cycle (PdPC) and to
compare it with allosteric cooperativity through both deterministic
(macroscopic) and stochastic (mesoscopic) models. The analysis
developed in the present paper indicates that the cooperativity in
the cyclic reaction is temporal, with energy ``stored'' in time
rather than in space as for allosteric cooperativity. This kind of
cooperativity utilizes multiple kinetic cycles in time, in contrast
to allosteric cooperativity that utilizes multiple subunits in a
protein.

It is necessary to emphasize that the essential similarities and
differences between temporal cooperativity and allosteric
cooperativity can only be put forward and discussed in stochastic
models.

In Section 2, we firstly introduce the deterministic and stochastic
model of the phosphorylation-dephosphorylation cycle. A
thermodynamic analysis confirms again the claim that the chemical
equilibrium state exists if and only if the phosphorylation
potential $\triangle G=0$; in this case the amplification of
sensitivity is completely abolished (Section 3).

In Section 4 and 5, we then provide comprehensive theoretical and
numerical analysis with the first-order and zero-order assumptions
in phosphorylation-dephosphorylation cycle respectively.

Furthermore, it is interestingly found in Section 6 that the
underlying mathematics of temporal cooperativity and allosteric
cooperativity are equivalent, and both of them can be expressed by
``dissociation constants'', which characterizes the essential
differences between the simple and ultrasensitive PdPC switches.
Nevertheless, the degree of allosteric cooperativity is restricted
by the total number of sites in a single enzyme molecule which can
not be freely regulated, while temporal cooperativity is only
restricted by the total number of molecules of the target protein
which can be regulated in a wide range and gives rise to the
ultrasensitivity phenomenon.

More implications of biochemistry are included in the discussion of
Section 7.

\section{Reversible kinetic model for covalent modification}

Many references  \mycite{GK,BQ2007,QH2003,SC,QH2007} have considered
the important phosphorylation-dephosphorylation cycle (PdPC)
catalyzed by kinase $E_1$ and phosphatase $E_2$, respectively. The
phosphorylation covalently modifies the protein $W$ to become $W^*$:
$$W+E_1+ATP\overset{a_1^0}{\underset{d_1}{\rightleftharpoons}}W\cdot E_1\cdot ATP\overset{k_1}{\underset{q_1^0}{\rightleftharpoons}}W^*+E_1+ADP;$$
$$W^*+E_2\overset{a_2}{\underset{d_2}{\rightleftharpoons}}W^*E_2\overset{k_2}{\underset{q_2^0}{\rightleftharpoons}}W+E_2+Pi.$$

Then at constant concentrations for $ATP$, $ADP$ and $Pi$,
introducing the pseudo reaction orders $a_1=a_1^0[ATP]$,
$q_1=q_1^0[ADP]$ and $q_2=q_2^0[Pi]$, these reactions become

$$Reaction~A1: W+E_1\overset{a_1}{\underset{d_1}{\rightleftharpoons}}WE_1;$$
$$Reaction~A2: WE_1\overset{k_1}{\underset{q_1}{\rightleftharpoons}}W^*+E_1;$$
$$Reaction~A3: W^*+E_2\overset{a_2}{\underset{d_2}{\rightleftharpoons}}W^*E_2;$$
$$Reaction~A4: W^*E_2\overset{k_2}{\underset{q_2}{\rightleftharpoons}}W+E_2.$$

This biochemical scheme is also isomorphic to another important
module in cellular signal transduction across the cell membrane,
namely the GTPase system.

From the chemical point of view, the total affinity \mycite{QH2003}
(intracellular phosphorylation potential) through the chemical
reactions $A1\rightarrow A2\rightarrow A3 \rightarrow A4$ is

\begin{eqnarray}
\triangle G&=&\triangle G_1+\triangle G_2+\triangle G_3+\triangle
G_4\nonumber\\
&=&k_BT\log\frac{a_1[W][E_1]}{d_1[WE_1]}+k_BT\log\frac{k_1[WE_1]}{q_1[W^*][E_1]}+k_BT\log\frac{a_2[W^*][E_2]}{d_2[W^*E_2]}+k_BT\log\frac{k_2[W^*E_2]}{q_2[W][E_2]}\nonumber\\
&=&k_BT\log\frac{a_1k_1a_2k_2}{d_1q_1d_2q_2}\\\nonumber
&=&k_BT\log\gamma,\nonumber
\end{eqnarray}
where $\gamma=\frac{a_1k_1a_2k_2}{d_1q_1d_2q_2}$ is called the
energy parameter.

Therefore, the system is in chemical equilibrium, if and only if
$\triangle G=0$, i.e.$\gamma=1$.

For the sake of sticking to the main point, the complete
deterministic and stochastic models as well as their thermodynamic
analysis are all put in the Appendix.

At the end of this subsection, it is indispensable to note that the
sustained high concentration of ATP ($\sim$1 mM) and low
concentrations of adenosine diphosphate (ADP) ($\sim$10 $\mu$M) and
Pi (orthophosphate) ($\sim$1 mM) give rise to an equilibrium
constant of $4.9\times 10^5$ M for ATP hydrolysis and the
phosphorylation potential in a normal cell is approximately 12 kcal
$mol^{-1}$ \mycite{Ho}.

\subsection{Reduced Mathematical models}

It is always supposed that the total concentration of $W$ and $W^*$
is much larger than that of the kinase and  phosphatase (i.e.
$W_T\gg E_{1T}+E_{2T}$ or equivalently $W_T=[W]+[W^*]$)
\mycite{GK,QH2003}, therefore, we can reasonably assume that the
time scale for the dynamics of enzymes $E_1$ and $E_2$ is much
faster than that for the dynamics of $W$ and $W^*$. Consequently,
the concentrations of $W$ and $W^*$
 can be recognized as constants when considering the dynamics of kinase $E_1$ and  phosphatase $E_2$, while
 the concentrations of $E_1$ and $E_2$ can be recognized as in
 steady states when considering the dynamics of $W$ and $W^*$.

Therefore, the dynamics of kinase and  phosphatase can be considered
separably:

\begin{eqnarray}\label{MM}
&&(a):W+E_1\overset{a_1}{\underset{d_1}{\rightleftharpoons}}WE_1\overset{k_1}{\underset{q_1}{\rightleftharpoons}}W^*+E_1\nonumber\\
&&(b):W+E_2\overset{q_2}{\underset{k_2}{\rightleftharpoons}}W^*E_2\overset{d_2}{\underset{a_2}{\rightleftharpoons}}W^*+E_2
\end{eqnarray}

The steady states in the above Michaelis-Menten kinetics has been
solved in the classic enzymology  \mycite{CB}, and the fluxes from
$W$ to $W^*$ and from $W^*$ to $W$ in reactions (a) and (b) of Eq.
\ref{MM} are
$$v_1([W])=\frac{\frac{V_1[W]}{K_1}}{1+\frac{[W]}{K_1}+\frac{[W^*]}{K_1^*}},
v_1^*([W^*])=\frac{\frac{V_1^*[W^*]}{K_1^*}}{1+\frac{[W]}{K_1}+\frac{[W^*]}{K_1^*}}$$
and
$$v_2([W])=\frac{\frac{V_2[W]}{K_2}}{1+\frac{[W]}{K_2}+\frac{[W^*]}{K_2^*}},
v_2^*([W^*])=\frac{\frac{V_2^*[W^*]}{K_2^*}}{1+\frac{[W]}{K_2}+\frac{[W^*]}{K_2^*}}$$
respectively, in which the parameters $V_1=k_1E_{1T}$,
$V_1^*=d_1E_{1T}$, $V_2=d_2E_{2T}$ and $V_2^*=k_2E_{2T}$ are the
maximal forward ($W\rightarrow W^*$) and backward ($W^*\rightarrow
W$) fluxes of the reactions (a) and (b); and
$K_1=\frac{d_1+k_1}{a_1}$, $K_2^*=\frac{d_2+k_2}{a_2}$,
$K_1^*=\frac{d_1+k_1}{q_1}$, $K_2=\frac{d_2+k_2}{q_2}$ are the
corresponding Michaelis constants.

Of more interest is the free energy constant
$$\gamma=\frac{a_1k_1a_2k_2}{d_1q_1d_2q_2}=\frac{V_1K_1^*V_2^*K_2}{V_1^*K_1V_2K_2^*}\equiv\frac{v_1([W])v_2^*([W^*])}{v_2([W])v_1^*([W^*])},$$
which doesn't vary with $[W]$ and $[W^*]$, and moreover makes the
model here not only more general but also more reasonable than the
semi-quantitative model introduced in \mycite{QH2003}.

\begin{figure}[h]
\centerline{\includegraphics[width=2.5in,height=3in,angle=270]{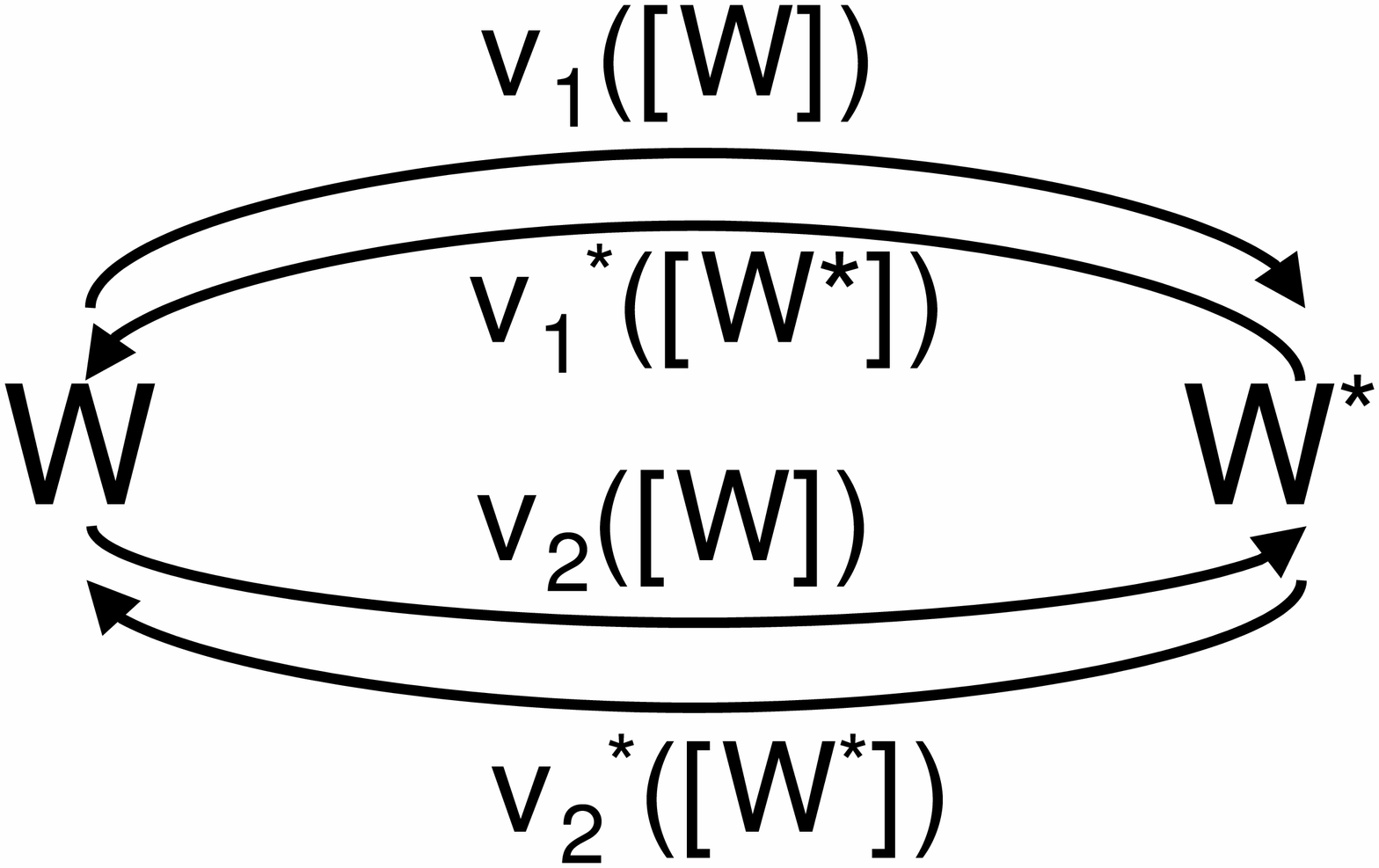}}
\caption[fig_reduced_model]{} \label{fig_reduced_model}
\end{figure}

Hence our model is now reduced the form of Fig.
\ref{fig_reduced_model}, which can be also found in the latest book
 \mycite{BQ2007} and reduced further to

\begin{equation}\label{Reduced_gen_model}
W\overset{f_1([W])}{\underset{f_2([W^*])}{\rightleftharpoons}W^*},
\end{equation}
where $f_1=v_1+v_2$ is the total flux from $W$ to $W^*$,
$f_2=v_1^*+v_2^*$ is the total flux from $W^*$ to $W$, and
$[W]+[W^*]=W_T$(constant).

\subsubsection{Deterministic model}\label{Sec_reduced-deter-model}

The ordinary differential equation of the model
(\ref{Reduced_gen_model}) is

\begin{equation}\label{Reduced-deter-model}
\frac{d[W^*]}{dt}=f_1(W_T-[W^*])-f_2([W^*]),
\end{equation}
whose steady state $[W^*]^{ss}$ satisfies
$f_1(W_T-[W^*]^{ss})=f_2([W^*]^{ss})$ and $[W]^{ss}=W_T-[W^*]^{ss}$.

What we concern most is the steady state fraction of phosphorylated
protein $W^*$, i.e. $\phi =\frac{[W^*]^{ss}}{W_T}$.

Beard and Hong Qian
  \mycite{BQ2007} have written down the general equation for $\phi =\frac{[W^*]^{ss}}{W_T}$ in the deterministic model under the restriction
 $W_T\gg E_{1T}+E_{2T}$ ($W_T=[W]+[W^*]$):

$$\theta=\frac{\mu\gamma[\mu-(\mu+1)\phi ](\phi -\frac{K_1^*(W_T+K_1)}{(K_1^*-K_1)W_T})K_2K_2^*(K_1^*-K_1)}{[\mu\gamma-(\mu\gamma+1)\phi ](\phi +\frac{K_2^*(W_T+K_2)}{(K_2-K_2^*)W_T})K_1K_1^*(K_2-K_2^*)},$$
where $\theta=\frac{V_1K_2^*}{K_1V_2^*}$,
$\mu=\frac{V_2K_2^*}{K_2V_2^*}$ and
$\gamma=\frac{V_1K_1^*V_2^*K_2}{V_1^*K_1V_2K_2^*}=\frac{a_1k_1a_2k_2}{d_1q_1d_2q_2}$.

In general chemical situation, we always have $K_1^*\gg K_1$,
$K_2\gg K_2^*$ºÍ$K_2\gg W_T$ (i.e. $q_1,q_2\ll 1$), then
$K_1^*-K_1\approx K_1^*$, $K_2-K_2^*\approx K_2$, $W_T+K_2\approx
K_2$, so the above equation can be simplified to
\begin{equation}\label{Gen_frac_eq}
\sigma\stackrel{def}{=}\frac{\theta
K_1}{K_2^*}=\frac{V_1}{V_2^*}=\frac{\mu\gamma[\mu-(\mu+1)\phi ](\phi
-1-\frac{K_1}{W_T})}{[\mu\gamma-(\mu\gamma+1)\phi ](\phi
+\frac{K_2^*}{W_T})}.
\end{equation}

If we let the free energy parameter $\gamma$ tends to infinity, then
$\mu=0$ (i.e. $q_1=q_2=0$). From (\ref{Gen_frac_eq}), one can get

$$\sigma=\frac{\phi (1-\phi +\frac{K_1}{W_T})}{(1-\phi )(\phi +\frac{K_2^*}{W_T})},$$
which is just the celebrated Goldbeter-Koshland equation
\mycite{GK} in their pioneer work on zero-order ultrasensitivity.

Solving the quadratic equation (\ref{Gen_frac_eq}), one can get that
\begin{equation}\label{Gen_f_star}
\phi =\frac{B-\sqrt{B^2-4AC}}{2A},
\end{equation}
where $A=1+\mu-\sigma(1+\frac{1}{\gamma\mu})$,
$B=\mu+(1+\mu)(1+\frac{K_1}{W_T})-\sigma[1-\frac{K_2^*}{W_T}(1+\frac{1}{\gamma\mu})]$,
and $C=\mu(1+\frac{K_1}{W_T})+\sigma\frac{K_2}{W_T}$. This
expression is put forward by Qian in  \mycite{QH2003}.

However, in such a deterministic model, the concentrations of
phosphorylated protein $W$ and its dephosphorylated state $W^*$ are
both the ensemble-averaged quantities, which can not really exhibit
the transition route between them and are unable to adequately
reveal the intrinsic essence of temporal cooperativity.

\subsubsection{Stochastic model: chemical master equation}

In order to illustrate the essence of temporal cooperativity, we
should turn to the stochastic model--chemical master equation. Let
$V$ be the volume of the system, then the total molecule number of
$W$ and $W^*$ is $N=W_TV$. Due to the existence of unavoidable
fluctuations, one can not determine the molecule numbers of each
species at any arbitrary time $t$, and instead can only determine
the probability that the vector representing the molecule numbers of
species $W$ and $W^*$ is $(N-i,i)$. According to
(\ref{Reduced_gen_model}), the chemical master equation model is
illustrated in Fig.  \ref{fig_Reduced_stoch}, where
$f_1(\frac{N-i}{V})V$ is the transition density jumping from state
$(N-i,i)$ to $(N-i-1,i+1)$, and $f_2(\frac{i}{V})V$ is the
transition density jumping from state  $(N-i,i)$ to $(N-i+1,i-1)$.

Similar transition models have recently emerged in
\mycite{QH2003,Elf2003}, but all of them did not explicitly insert
the volume parameter $V$ into their models, ignoring the variety of
stochasticity related with the volume.

Denote the probability of the state $(N-i,i)$ at time $t$ as
$P(N-i,i;t)$, then it satisfies the {\bf chemical master equation}
\begin{eqnarray}\label{Reduced-stoch-model}
\frac{dP(N,0;t)}{dt}&=&f_2(1/V)VP(N-1,1;t)-f_1(N/V)VP(N,0;t);\nonumber\\
\frac{dP(N-i,i;t)}{dt}&=&f_1((N+1-i)/V)VP(N+1-i,i-1;t)\nonumber\\
&&+f_2((i+1)/V)VP(N-1-i,i+1;t)\nonumber\\
&&-[f_1((N-i)/V)+f_2(i/V)]VP(N-i,i;t),~i=1,2,\cdots,N-1;\nonumber\\
\frac{dP(0,N;t)}{dt}&=&f_1(1/V)VP(1,N-1;t)-f_2(N/V)VP(0,N;t).
\end{eqnarray}

In probability theory, such a random-walk model is called the
one-dimensional birth-and-death process, which is a special Markov
chain, and the above equation is called the Kolmogorov forward
equation (also called Fokker-Planck equation) of the continuous-time
Markov chain with transition density matrix $Q=(q_{\xi\eta})$, in
which
$$q_{\xi\eta}=\left\{\begin{array}{ll}f_1(\frac{N-i}{V})V,&\xi=(N-i,i),\eta=(N-i-1,i+1),i=0,1,\cdots,N-1;\\f_2(\frac{i}{V})V,&\xi=(N-i,i),\eta=(N-i+1,i-1),i=1,\cdots,N;\\-f_1(\frac{N-i}{V})V-f_2(\frac{i}{V})V,&\xi=\eta=(N-i,i),i=1,2,\cdots,N-1;\\-f_1(\frac{N}{V})V,&\xi=\eta=(N,0);\\-f_2(\frac{N}{V})V,&\xi=\eta=(0,N);\\0,&else.\end{array}\right.$$

Two points of importance are worth emphasizing: (i) there is a basic
assumption for the validity of this reduced stochastic model
(\ref{Reduced-stoch-model}), that is the time scale for the dynamics
of enzymes $E_1$ and $E_2$ are much faster than that for the
dynamics of $W$ and $W^*$ so as to ensure the Markovian property of
this simplified model, especially when the functions $f_1$ and $f_2$
are nonlinear; (ii) According to the Kolmogorov cyclic condition
(See Appendix), since there is no cycles consisting of more than two
states in the chemical master equation model (Fig.
\ref{fig_Reduced_stoch}), detailed balance condition is satisfied.
However, the detailed balance condition of this reduced stochastic
model does not allude to the chemical equilibrium state of the
original model, because the reversibility(equilibrium) of the
complete model (i.e. $\gamma=1$) is not equivalent to the
reversibility of this reduced model.

From (\ref{Reduced-stoch-model}), in the steady state, the ratio of
the probabilities of the states  $(N-i,i)$ and $(N,0)$ is
$\Pi_{j=1}^i\left[\frac{f_1((N+1-j)/V)V}{f_2(j/V)V}\right]=\Pi_{j=1}^i\left[\frac{f_1((N+1-j)/V)}{f_2(j/V)}\right]$
(See Appendix for derivation), then the steady distribution of the
state $(N-i,i)$ is
\begin{eqnarray}\label{Gen-stoch-dist}
P^{ss}(N-i,i)=\frac{\Pi_{j=1}^i\frac{f_1((N+1-j)/V)}{f_2(j/V)}}{1+\sum_{i=1}^N\Pi_{j=1}^i\frac{f_1((N+1-j)/V)}{f_2(j/V)}},
\end{eqnarray}
and the averaged molecule number of $W^*$ is
$$\langle W^* \rangle=\frac{\sum_{i=1}^Ni\Pi_{j=1}^i\frac{f_1((N+1-j)/V)}{f_2(j/V)}}{1+\sum_{i=1}^N\Pi_{j=1}^i\frac{f_1((N+1-j)/V)}{f_2(j/V)}}.$$

Similar to the deterministic model, we introduce the ratio of the
averaged molecule number $\langle W^* \rangle$ of phosphorylated
protein molecules and the total molecule number $N$,

\begin{equation}\label{f_star_star}
\langle \phi \rangle \stackrel{def}{=}\frac{\langle W^*
\rangle}{N}=\frac{\sum_{i=1}^Ni\Pi_{j=1}^i\frac{f_1((N+1-j)/V)}{f_2(j/V)}}{N(1+\sum_{i=1}^N\Pi_{j=1}^i\frac{f_1((N+1-j)/V)}{f_2(j/V)})}.
\end{equation}

When $N=W_TV$, it is easy to find that if $j>[W^*]^{ss}V$, then
$\frac{f_1((N+1-j)/V)}{f_2(j/V)}<1$, and if $j<[W^*]^{ss}V$, then
$\frac{f_1((N+1-j)/V)}{f_2(j/V)}>1$. Hence, the state with the
highest probability is $(W_TV-[W^*]^{ss}V,[W^*]^{ss}V)$. When
$V\rightarrow\infty$, all the probability will tend to centralize on
the state $((W_T-[W^*]^{ss})V,[W^*]^{ss}V)$, which perfectly
corresponds to the steady state of the deterministic model by the
mathematical theory of T. Kurtz  \mycite{Kur}. Consequently, we have
$\langle \phi \rangle \rightarrow \phi $, when $V\rightarrow\infty$.

In this stochastic model, $\langle \phi \rangle $ does not usually
have a simple explicit expression, but it will be shown in the
following sections, under different reasonable approximations that
correspond to the simple and ultrasensitive PdPC switches
respectively, the expression of $\langle \phi \rangle $ is then
clear and definite.

In addition, due to the nonlinearity of functions $f_1$ and $f_2$,
although $\langle \phi \rangle \rightarrow \phi $ when the molecule
numbers tend to infinity, the graph of $\langle \phi \rangle $ is
more gradual than that of $\phi $, which is pointed out recently by
Berg, et al. \mycite{BPE}.

\subsubsection{Dissociation constants}

If the functions $f_1$, $f_2$ are both linear, i.e.
$f_1([W])=\alpha_1[W]$, and $f_2([W^*])=\alpha_2[W^*]$. In this
case, according to (\ref{Reduced-deter-model}) and
(\ref{f_star_star}), it is derived that $\langle \phi \rangle =\phi
=\frac{\alpha_1/\alpha_2}{1+\alpha_1/\alpha_2}$(hyperbolic)
illustrating no cooperative effect.

In order to estimate the degree of cooperative phenomenon in the
PdPC switch, we introduce the dissociation constants similar to the
Adair constants \mycite{CB} in the allosteric cooperative
phenomenon.

For the state $(N+1-j,j-1)$, there have already been $(j-1)$
molecules transited from $W$ to $W^*$, thus there are $(N+1-j)$ ways
of transiting for the next molecule of $W$ to $W^*$. Similarly, for
the state $(N-j,j)$, there have already been $j$ molecules transited
from $W$ to $W^*$, and there are $j$ ways of transiting for the next
molecule of $W^*$ back to $W$.

Define quantities $K_j=\frac{(N+1-j)f_2(j/V)}{jf_1((N+1-j)/V)}$,
representing the ``dissociation capability'' of the j-th molecule in
the state $(N-j,j)$ transiting back from the activated species $W^*$
to the inactivated one $W$, which are called ``dissociation
constants'', and their reciprocals are representing the
``association capability'' of the j-th molecule transiting from the
inactivated species $W$ to the activated one $W^*$, which can be
called ``association constants''.

In Section 6, we will show that the underlying mathematics of
temporal cooperativity and allosteric cooperativity are equivalent,
and both of them can be expressed by ``dissociation constants'',
which reveals the essential differences between the simple and
ultrasensitive PdPC switches. So here it is worth rewriting the
formula (\ref{f_star_star}) by the dissociation constants as
$$\langle \phi \rangle=\frac{\sum_{i=1}^N \frac{(N-1)!}{(i-1)!(N-i)!}\frac{1}{\prod_{j=1}^i K_j}}{1+\sum_{i=1}^N \frac{N!}{i!(N-i)!}\frac{1}{\prod_{j=1}^i K_j}},$$
which is essentially same as the general Adair scheme of allosteric
cooperativity (\ref{Adair-scheme}).

With these in our model, there exists the temporal cooperative
phenomenon if the quantities $\{K_j,~j=1,2,\cdots,N\}$ successively
decreases, which means the more number of molecules of $W^*$ is, the
larger the association constant of the next molecule transiting from
the state $W$ to $W^*$ becomes. Furthermore, the cooperative
phenomenon appears more and more distinct when the gradient of the
decreasing quantities $\{K_j,~j=1,2,\cdots,N\}$ increases.

\section{Chemical equilibrium state ($\gamma=1$): no switch}

Sensitivity amplification requires energy consumption, and
phosphorylation potential can be used to improve specificity in
biomolecular recognition and robustness in cell development
\mycite{QH2003}.

\subsection{In the deterministic model}

When $\gamma=1$, this system is in chemical equilibrium state and we
have
$\frac{f_1([W])}{f_2([W^*])}=\frac{v_1([W])}{v^{*}_1([W^*])}=\frac{v_2([W])}{v^{*}_2([W^*])}=\mu\frac{[W]}{[W^*]}$,
recalling $\mu=\frac{d_2q_2}{a_2k_2}$ is a constant. Hence, by
(\ref{Gen_f_star}) $\phi=\frac{\mu}{\mu+1}$ is a constant, which
does not vary with the concentrations of the kinase and phosphatase,
and implies that there is no biological switch here.

It is necessary to point out that in the simplified equation
(\ref{Gen_frac_eq}), if $\gamma=1$ and $\phi \neq
\frac{\mu}{\mu+1}$, then by canceling a nonzero factor
$$\frac{V_1}{V_2^*}=\frac{\mu(\phi -1-\frac{K_1}{W_T})}{\phi +\frac{K_2^*}{W_T}},$$
but the right side is negative since $\phi$ is less than 1, which
contradicts the left side. Therefore, this simplified equation still
preserves the fact that the PdPC switch is a nonequilibrium
phenomenon ($\gamma\neq 1$), which confirms the significant belief
that biological amplification needs energy.

\subsection{In the stochastic model}

The model discussed in \mycite{QH2003} is deterministic. It will be
shown here that the same conclusion also holds in the stochastic
model.

Since if $\gamma=1$, then
$\frac{f_1([W])}{f_2([W^*])}=\frac{v_1([W])}{v^{*}_1([W^*])}=\frac{v_2([W])}{v^{*}_2([W^*])}=\mu\frac{[W]}{[W^*]}$,
and the steady distribution of the state $(N-i,i)$ is
$\frac{N!}{i!(N-i)!}\mu^i/(1+\mu)^N$(Binomial distribution), so
$$\langle \phi \rangle =\frac{\langle W^* \rangle}{N}=\frac{\sum_{i=1}^Ni\frac{N!}{i!(N-i)!}\mu^i}{N(1+\sum_{i=1}^N\frac{N!}{i!(N-i)!}\mu^i)}=\frac{\mu}{1+\mu},$$
which is the same as the quantity $\phi $ in the deterministic model
and also implies that the amplification of sensitivity is completely
abolished.

Furthermore, the dissociation constants $\{K_i, 1\leq i\leq N\}$ are
all equal to $\frac{1}{\mu}$, unaltering with the concentrations of
kinase and phosphatase.

\section{Simple PdPC switch ($\gamma\neq 1$)}

\subsection{Theoretical analysis of the first-order linear
approximation (i.e. $f_1$ and $f_2$ are linear)}

Suppose $W_T\ll K_1,K_2^*\ll K_1^*, K_2$(non-saturated ), then

$$f_1([W])=v_1([W])+v_2([W])=\frac{\frac{V_1[W]}{K_1}}{1+\frac{[W]}{K_1}+\frac{[W^*]}{K_1^*}}+\frac{\frac{V_2[W]}{K_2}}{1+\frac{[W]}{K_2}+\frac{[W^*]}{K_2^*}}\approx \frac{V_1[W]}{K_1}+\frac{V_2[W]}{K_2},$$
and
$$f_2([W^*])=v_1^*([W^*])+v_2^*([W^*])=\frac{\frac{V^*_1[W^*]}{K^*_1}}{1+\frac{[W]}{K_1}+\frac{[W^*]}{K_1^*}}+\frac{\frac{V^*_2[W^*]}{K^*_2}}{1+\frac{[W]}{K_2}+\frac{[W^*]}{K_2^*}}\approx \frac{V_2^*[W^*]}{K_2^*}+\frac{V_1^*[W^*]}{K_1^*},$$
are both first-order, which is just the ordinary PdPC switch
discussed in
 \mycite{QH2007}.

The steady state of the deterministic model is
$[W]^{ss}=\frac{W_T}{1+\alpha}$ and
$[W^*]^{ss}=\frac{W_T\alpha}{1+\alpha}$, where
$\alpha=\frac{\frac{V_1}{K_1}+\frac{V_2}{K_2}}{\frac{V_2^*}{K_2^*}+\frac{V_1^*}{K_1^*}}$.
And since $\phi -1-\frac{K_1}{W_T}\approx -\frac{K_1}{W_T}$ and
$\phi +\frac{K_2^*}{W_T}\approx \frac{K_2^*}{W_T}$, the equation
(\ref{Gen_frac_eq}) is reduced to
$$\theta=\frac{V_1 K_2^*}{V_2^* K_1}=\frac{\mu\gamma[(\mu+1)\phi -\mu]}{[\mu\gamma-(\mu\gamma+1)\phi ]}.$$
i.e.
$$\phi =\frac{\theta+\mu}{\theta+\mu+\theta/(\mu\gamma)+1}=\frac{\alpha}{1+\alpha}.$$

And in the stochastic model, the steady distribution of the state
$(N-i,i)$ is (from (\ref{Gen-stoch-dist}))
$\frac{N!}{i!(N-i)!}\alpha^i/(1+\alpha)^N$(Binomial distribution),
then
$$\langle \phi \rangle =\frac{\langle W^* \rangle}{N}=\frac{\sum_{i=1}^Ni\frac{N!}{i!(N-i)!}\alpha^i}{N(1+\sum_{i=1}^N\frac{N!}{i!(N-i)!}\alpha^i)}=\frac{\alpha}{1+\alpha},$$
which is the same as the quantity $\phi $ in the deterministic
model.

Furthermore,
$\alpha=\frac{\frac{V_1}{K_1}+\frac{V_2}{K_2}}{\frac{V_2^*}{K_2^*}+\frac{V_1^*}{K_1^*}}$
is an increasing hyperbolic function of $E_{1T}$. So $\langle \phi
\rangle =\phi $ is also an increasing hyperbolic function of
$E_{1T}$ illustrating no cooperative effect either, which implies
that the $N$ molecules of $W$ and $W^*$ are all independent.

In the real organism, the signal molecule is the kinase $E_1$, so
one should use the total concentration $E_{1T}$ of $E_1$ as the
control parameter rather than $\theta=\frac{V_1K_2^*}{V^*_2K_1}$
used in
 \mycite{BQ2007}.

The variance of the molecule number of $W^*$ is
$\Sigma=\frac{\alpha}{(1+\alpha)^2}W_TV$, so the relative standard
error is $\frac{\sqrt{\Sigma}}{\phi V}=\sqrt{\frac{W_T}{\alpha
V}}\rightarrow 0$ when $V\rightarrow \infty$, according to the
mathematical theory of Kurtz  \mycite{Kur}.

\subsection{Numerical verification by simulation}



Now, we could numerically analyze the cooperative effect in this
simple PdPC switch.

Fig.  \ref{Fig_simple_switch_fig2_1} illustrates the curve of $\phi
$ with respect to $E_{1T}$ based on the formula (\ref{Gen_f_star})
of the deterministic model (\ref{Reduced-deter-model}) of the simple
PdPC switch without the first-order linear approximation. It
presents a simple hyperbolic curve, implying non-cooperative effect.

Fig.  \ref{Fig_simple_switch_fig2} illustrates the curves of
$\langle \phi \rangle $ with respect to $E_{1T}$ in the stochastic
model (\ref{Reduced-stoch-model}) of the simple PdPC switch without
the first-order linear approximation at different volumes, all of
which also presents the simple hyperbolic shape.

Fig.  \ref{Fig_simple_switch_coop} represents the dissociation
constants $\{K_i\}$ of temporal cooperativity with different
volumes. It is found that in such a simple PdPC switch, these
dissociation constants are all very close to $1$ regardless of the
variety of volumes, reconfirming no obvious cooperative phenomenon.

\section{Ultrasensitive PdPC switch}

\subsection{Theoretical analysis of the zero-order approximation}

Suppose $K_2,K_1^*\gg W_T\gg K_1,K_2^*$(saturated), and $K_2^*\ll
K_2$, $K_1\ll K_1^*$, one can arrive at the limit case (
$\frac{[W^*]}{K_1^*}\approx 0$ and $\frac{[W]}{K_2}\approx 0$)

$$f_1([W])=v_1([W])+v_2([W])=\frac{\frac{V_1[W]}{K_1}}{1+\frac{[W]}{K_1}+\frac{[W^*]}{K_1^*}}+\frac{\frac{V_2[W]}{K_2}}{1+\frac{[W]}{K_2}+\frac{[W^*]}{K_2^*}}\approx V_1,$$
and
$$f_2([W^*])=v_1^*([W^*])+v_2^*([W^*])=\frac{\frac{V^*_1[W^*]}{K^*_1}}{1+\frac{[W]}{K_1}+\frac{[W^*]}{K_1^*}}+\frac{\frac{V^*_2[W^*]}{K^*_2}}{1+\frac{[W]}{K_2}+\frac{[W^*]}{K_2^*}}\approx V^*_2.$$

These are both in zeroth order case, which should be considered as
nonlinear since $f_1(0)\neq 0$ and $f_2(0)\neq 0$. This is just the
situations of ultrasensitive PdPC switch discussed in
 \mycite{QH2003} and zero-order ultrasensitivity phenomenon put
forward by Goldbeter and Koshland  \mycite{GK}. The Hill coefficient
of the response curve can approach thousands and tens of thousands.
It is worth pointing out that such a limit case can only be achieved
when $\gamma\neq 1$, since otherwise $V_1^*V_2\gg V_1V_2^*$ which
contradicts the zero-order approximation.

In the deterministic model of this limit case, we have $\phi
=\delta_{\{V_1>V_2^*\}}$, which is a step function with ideal
infinite sensitivity. And in the stochastic model, the steady
distribution of the state $(N-i,i)$ is
$\frac{\alpha^i}{N(1+\sum_{i=1}^N\alpha^i)}$(truncated geometric
distribution), so

\begin{equation}
\langle \phi \rangle =\frac{\langle W^*
\rangle}{N}=\frac{\sum_{i=1}^Ni\alpha^i}{N(1+\sum_{i=1}^N\alpha^i)}=\left\{\begin{array}{ll}\frac{N\alpha^{N+1}-\frac{\alpha^{N+1}-\alpha}{\alpha-1}}{N(\alpha^{N+1}-1)}&
\alpha\neq 1;\\1/2&\alpha=1,\end{array}\right.
\end{equation}
where $\alpha=\frac{V_1}{V_2^*}$ is the ratio of the forward flux
from $W$ to $W^*$ and the backward flux from $W^*$ to $W$.

Obviously, $\langle \phi \rangle $ is an increasing function of
$\alpha$, and consequently an increasing function of $E_{1T}$. And
when $N\rightarrow\infty$, one has $\langle \phi \rangle \rightarrow
1$, if $\alpha>1$; $\langle \phi \rangle \rightarrow 0$, if
$\alpha<1$ (See Fig.  \ref{Fig_ultra_switch_fig2}). The classical
Hill coefficient in this case $n_H=2\frac{d\log \langle \phi \rangle
}{d\log \alpha}|_{\langle \phi
\rangle=\frac{1}{2}}=\frac{1}{3}N+\frac{2}{3}$. Therefore, when the
total molecule number $N$ tends to infinity, the Hill coefficient
can increase to an arbitrary value.

Hence, when the Michaelis constants $K_1,K_2$ are quite small, the
ultrasensitive cooperative phenomenon emerges both in deterministic
and stochastic models, although their sensitivities can not be as
high as in the limit case discussed above.

\subsection{Numerical verification by simulation}

Firstly, we investigate the cooperative phenomenon in the limit case
of zero-order approximation.


Fig.  \ref{Fig_ultra_switch_fig2} illustrates the curves of $\langle
\phi \rangle $ with respect to $E_{1T}$  at different volumes in the
stochastic model of ultrasensitive PdPC switch under the zero-order
approximation, in which it is found that the sensitivities of these
curves are increasing with the volumes (molecule numbers) and
finally approaches the ideal jumping curve of $\phi$ with infinite
sensitivity.

Secondly, we turn to discuss the cooperative phenomenon without the
zero-order approximation.

Fig. \ref{Fig_ultra_switch_fig3_1} illustrates the curve of $\phi $
with respect to $E_{1T}$ based on the equation (\ref{Gen_f_star}) in
the deterministic model (\ref{Reduced-deter-model}) of the
ultrasensitive PdPC switch without the zero-order approximation,
whose sensitivity is less than that in Fig.
\ref{Fig_ultra_switch_fig2} but much larger than that in Fig.
\ref{Fig_simple_switch_fig2}.

Fig. \ref{Fig_ultra_switch_fig3} illustrates the curves of $\langle
\phi \rangle $ with respect to $E_{1T}$ at different volumes in the
stochastic model (\ref{Reduced-stoch-model}) by formula
(\ref{f_star_star}) of the ultrasensitive PdPC switch without the
zero-order approximation, in which it is found that the
sensitivities of these curves are increasing with the
volumes(molecule numbers).

There is a significant difference in terms of the Hill coefficient
(degree of steepness) between the zeroth order approximate solution
(Fig. \ref{Fig_ultra_switch_fig2}) and the exact solution (Fig.
\ref{Fig_ultra_switch_fig3}). Although the trend in both cases is
the same, namely larger molecule numbers gives more cooperativity,
the latter one clearly approaches a limit, which is just the curve
of $\phi $  with finite sensitivity in Fig.
\ref{Fig_ultra_switch_fig3_1}. This accords well with the famous
mathematical theory of T.G. Kurtz \mycite{Kur}, which says the
deterministic model is just the infinite volume limit of the
chemical master equation as the concentration parameters are
unaltered.

Fig. \ref{Fig_ultra_switch_coop} represents the dissociation
constants $\{K_i\}$ of cooperativity with different volumes. It is
found that in the ultrasensitive PdPC switch, these dissociation
constants clearly decrease, and the gradient increases with the
total molecule numbers, suggesting more and more distinct
cooperative phenomenon.

\section{Mathematical equivalence to allosteric cooperativity}\label{sec-allos-coop}

In this section, we will investigate the equivalence of the
underlying mathematics in temporal cooperativity and allosteric
cooperativity, both of which can be expressed by ``dissociation
constants'', which also raises the essential differences between the
simple and ultrasensitive PdPC switches (Fig.
\ref{Fig_simple_switch_coop} and Fig. \ref{Fig_ultra_switch_coop}).

Fig.  \ref{fig_allos_model} is the general model of allosteric
cooperative phenomenon including both the famous MCJ and KNF models
\mycite{MCJ,KNF}, which can all be expressed by the Adair scheme,
first proposed by Adair \mycite{Ad} in relation to the binding of
oxygen to haemoglobin. In this model, the concentration of the
substrate $S$ is fixed, and the vector $(N-i,i)$ represents the
state in which there are $i$ sites occupied with substrates among
the total $N$ sites.

\begin{figure}[h]
\centerline{\includegraphics[width=2in,height=4in,angle=270]{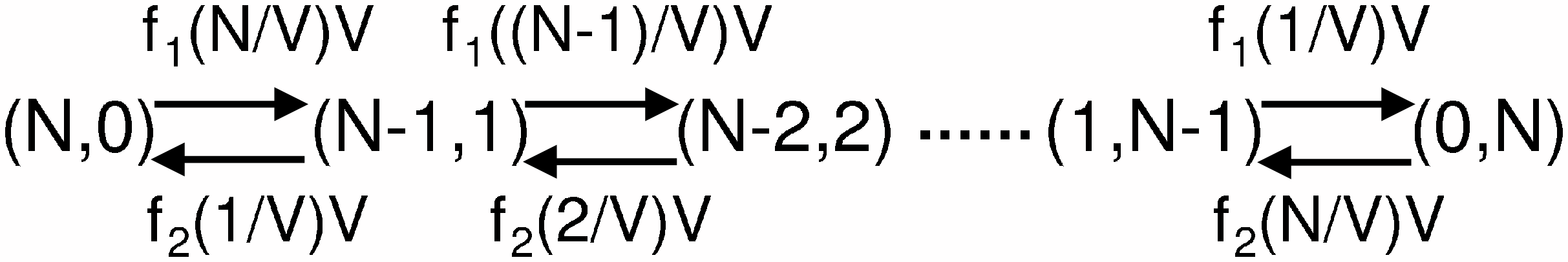}}
\caption[fig_Reduced_stoch]{} \label{fig_Reduced_stoch}
\end{figure}

\begin{figure}[h]
\centerline{\includegraphics[width=4in,height=3in]{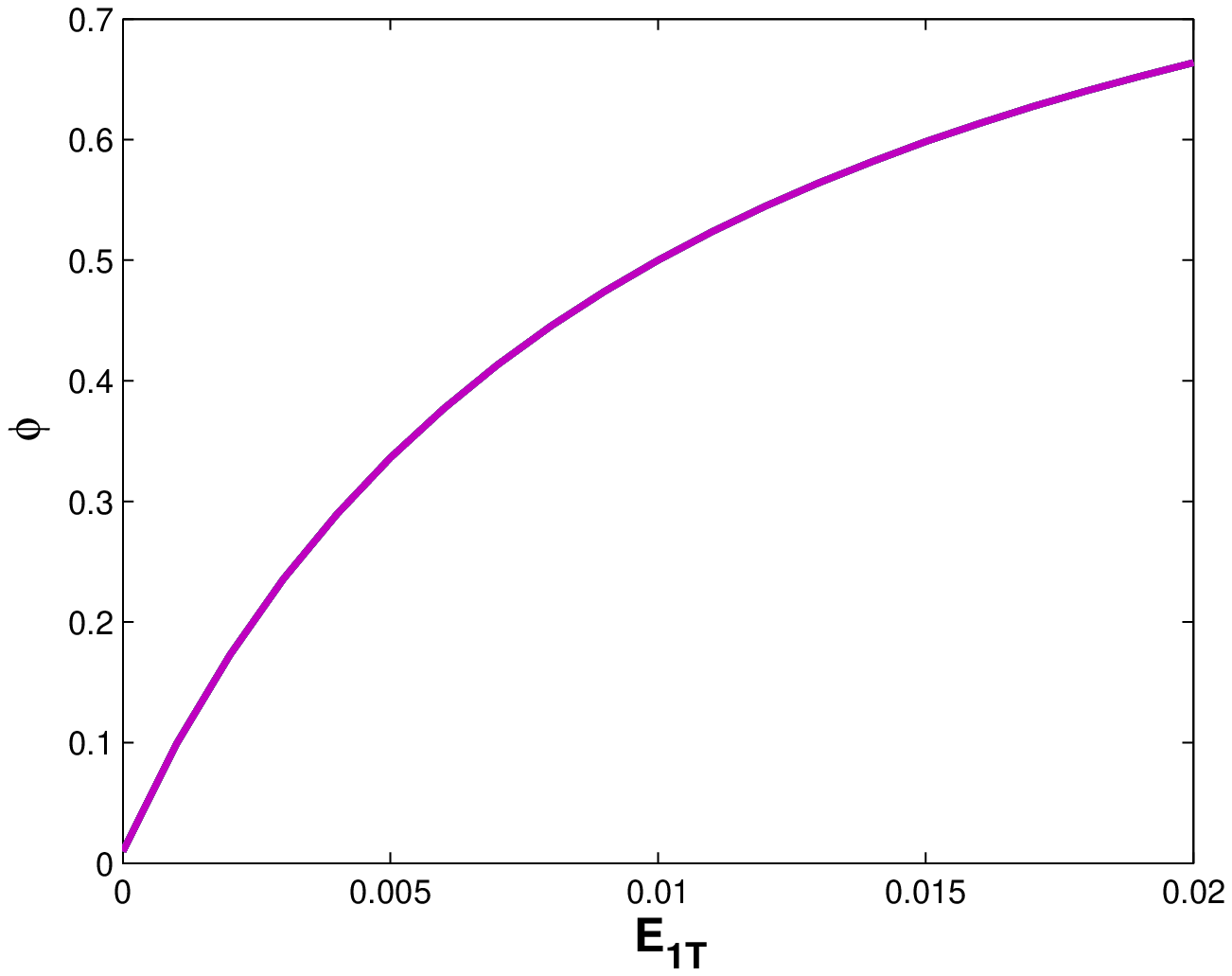}}
\caption[Fig_simple_switch_fig2_1]{}
\label{Fig_simple_switch_fig2_1}
\end{figure}

\begin{figure}[h]
\centerline{\includegraphics[width=4in,height=3in]{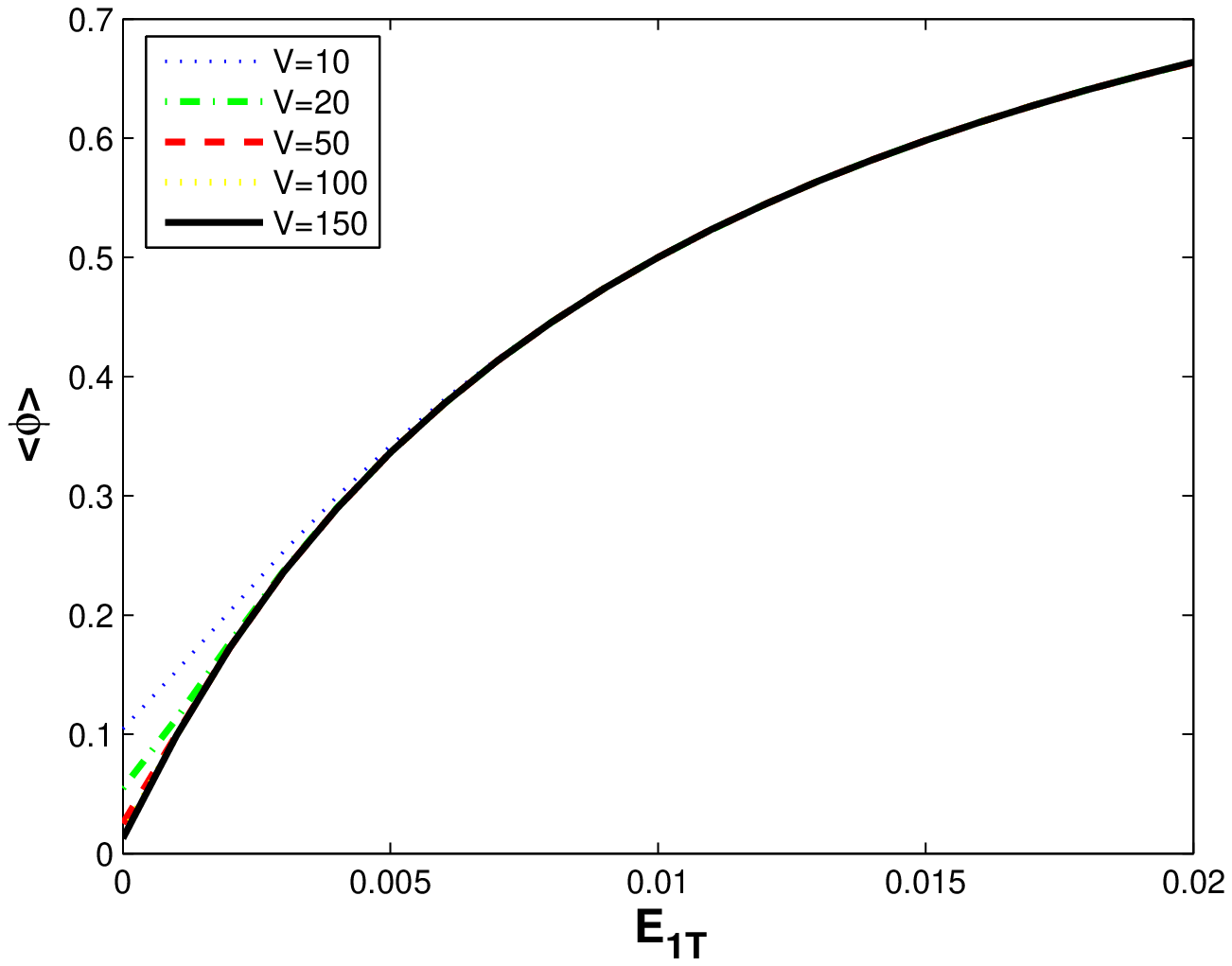}}
\caption[Fig_simple_switch_fig2]{} \label{Fig_simple_switch_fig2}
\end{figure}

It is very important to point out that Fig.  \ref{fig_allos_model}
is nearly the same as Fig.  \ref{fig_Reduced_stoch}, where the
temporal cooperativity is on the scale of the $N$ sequential
phosphorylation-dephosphorylation cycles. The sequential states in
Fig. \ref{fig_Reduced_stoch} are adjacent in time rather than in
space which is the case in allosteric cooperativity. The model in
Fig. \ref{fig_allos_model} is a special case of the model in Fig.
\ref{fig_Reduced_stoch} when
$\frac{f_1(N+1-i/V)}{f_2(i/V)}=\frac{(N+1-i)[S]}{iK_i}$.

Meanwhile, a similar model to (\ref{Reduced_gen_model}) can also be
written down as
\begin{equation}\label{Gen_coop_model}
E\overset{f_1(n_E)}{\underset{f_2(n_{E^*})}{\rightleftharpoons}E^*},
\end{equation}
where $E$ and $E^*$ represent the unoccupied and occupied states of
single site respectively; $n_E$ and $n_{E^*}$ are the numbers of
unoccupied and occupied sites respectively. Hence, $n_E+n_{E^*}=N$,
and (\ref{Gen_coop_model}) is equivalent to the model
(\ref{Reduced_gen_model}) as long as the key equality
$\frac{f_1(N+1-n_{E^*})}{f_2(n_{E^*})}=\frac{(N+1-n_{E^*})[S]}{n_{E^*}K(n_{E^*})}$
holds, where $K(n_{E^*})$ is the dissociation constant of the
$n_{E^*}$-th molecule of the substrate.

These two kinds of cooperativity phenomena both come from the
nonlinearity of functions $f_1$ and $f_2$ (i.e. the varying of
$K_i$), but the former emerges from the complex chemical reactions
while the latter arises from the allosteric interactions between
different sites. Actually, although there is no direct interaction
between the substrate enzymes, the total $N$ molecules of $W$ and
$W^*$ are not really independent: they all compete for the single
kinase and phosphatase and hence there are implicit interactions
between them. Because this interaction is not through space, but
instead is sequential in time, so Hong Qian
\mycite{QH2003,QH_BJREV2008} refer to it as temporal cooperativity.

Moreover, the meanings of the quantity $N$ in Fig.
\ref{fig_Reduced_stoch} and Fig.  \ref{fig_allos_model} are totally
different: the former represents the total molecule number in the
temporal cooperativity model and the latter represents the total
number of sites on a single enzyme molecule respectively. Hence, the
degree of allosteric cooperativity is restricted by the total number
of sites in a single enzyme molecule which can not be very high (see
(\ref{Max_hill})) and freely regulated, while temporal cooperativity
is only restricted by the total molecule number of the target
protein which can be regulated in a wide range and gives rise to the
ultrasensitivity phenomenon.

In order to be consistent with the previous sections, we still use
the symbol $\phi $ here to represent the fractional saturation.

Cooperativity can be generally considered in relation to the Adair
scheme, and the general form of Adair equation is

\begin{equation}\label{Adair-scheme}
\phi =\frac{\sum_{i=1}^N
\frac{(N-1)!}{(i-1)!(N-i)!}\frac{c^i}{\prod_{j=1}^i
K_j}}{1+\sum_{i=1}^N \frac{N!}{i!(N-i)!}\frac{c^i}{\prod_{j=1}^i
K_j}},\nonumber
\end{equation}
where $c=[S]$, $K_j=\frac{(N-j+1)c[ES_{j-1}]}{j[ES_{j}]}$ is the
dissociation constant of the $j-th$ molecule of the substrate
(regardless of site).

Consequently, there is an important corollary, that is the Hill
coefficient of the $[S]-\phi $ curve determined by the Adair
equation can not exceed the total number $N$ of sites on a single
enzyme, i.e.

\begin{eqnarray}\label{Max_hill}
n_H&=&2\frac{d\log\phi }{d\log c}|_{\phi =\frac{1}{2}}\nonumber\\
&=&[4\frac{\sum_{i=1}^N
i\frac{(N-1)!}{(i-1)!(N-i)!}\frac{c^i}{\prod_{j=1}^i
K_j}}{1+\sum_{i=1}^N \frac{N!}{i!(N-i)!}\frac{c^i}{\prod_{j=1}^i
K_j}}-4N(\phi )^2]|_{\phi =\frac{1}{2}}\nonumber\\
&\leq&[4N\phi -4N(\phi )^2]|_{\phi =\frac{1}{2}}=N.
\end{eqnarray}

It is thought that
 \mycite{CB}``any valid equation to describe binding of a
ligand to a micromolecule at equilibrium must be''Adair equation,
and in many cases, the Adair constants can be actually regarded as
``statistical factors'' when fitting experimental data.

In addition, the definition of cooperativity in relation to the
Adair constants and the Hill plot are not equivalent, and they do
not always result in the same sign of cooperativity. However, in
several simple cases there is good agreement between them
 \mycite{RC1987}.

For instance, when $N=2$, $\phi
=\frac{\frac{c}{K_1}+\frac{c^2}{K_1K_2}}{1+\frac{2c}{K_1}+\frac{c^2}{K_1K_2}}$£¬
then $\frac{d\phi }{d\log
c}=\frac{\frac{c}{K_1}+\frac{2c^2}{K_1K_2}+\frac{c^3}{K_1^2K_2}}{(1+\frac{2c}{K_1}+\frac{c^2}{K_1K_2})^2}$,
and when $\phi =\frac{1}{2}$, the half saturation concentration
$K_{0.5}=\sqrt{K_1K_2}$. So the Hill coefficient
$n_H=\frac{2}{(1+\sqrt{\frac{K_2}{K_1}})}$. Hence, $n>1$ is
equivalent to $K_2<K_1$, and $n<1$ is equivalent to $K_2>K_1$.

In the subsections below, we will briefly review several famous
examples, and our aim is to uniformly describe the allosteric
cooperative phenomenon by the Adair scheme so that to compare with
the temporal cooperative phenomenon (See Table \ref{tab1} in this
section).

\subsection{Symmetric model}

Monod, Changeux and Jacob  \mycite{MCJ}  studied many examples of
cooperative and allosteric phenomenon, and concluded that they were
closely related and that conformational flexibility probably
contributed for both. Subsequently, Monod, Wyman and Changeux
 \mycite{MWC} proposed a general symmetric model to explain both phenomena, which requires each site can exist in two
different conformations, $R$ and $T$, and all sites must be in the
same conformation.

\subsubsection{Two sites}

Symmetric model for a two-site protein is illustrated in Fig.
\ref{fig5}, where $A$ is the substrate. This example is from
\mycite{CB}.

$L$ is the equilibrium constant between the two conformations. $K_R$
and $K_T$ are the dissociation constants of the two conformations
$R$ and $T$ bound with the substrate $A$. The fractional saturation
takes the following form

$$\phi =\frac{[R_2A]+2[R_2A_2]+[T_2A]+2[T_2A_2]}{2([R_2]+[R_2A]+[R_2A_2]+[T_2]+[T_2A]+[T_2A_2])},$$

Furthermore,

$$\phi =\frac{[A]/K_R+[A]^2/K_R^2+L[A]/K_T+L[A]^2/K_T^2}{(1+[A]/K_R)^2+L(1+[A]/K_T)^2},$$
which can be rearranged into the form of the Adair equation

$$\phi =\frac{\frac{[A]}{K_1}+\frac{[A]^2}{K_1K_2}}{1+2\frac{[A]}{K_1}+\frac{[A]^2}{K_1K_2}},$$
where the dissociation constants $K_1=\frac{1+L}{1/K_R+L/K_T}$, and
$K_2=\frac{1/K_R+L/K_T}{1/K_R^2+L/K_T^2}$.

According to the Cauchy inequality, one has $K_1\geq K_2$, which
implies positive cooperative phenomenon. Moreover, $K_1> K_2$ is
equivalent to the condition that $0<L<\infty$ and $K_R\neq K_T$.

\subsubsection{$N$ sites}

Straightforward generalizing the results above to the case of $N$
sites, one has
$\frac{[R_NA_i]}{[R_N]}=\frac{N!}{i!(N-i)!}[A]^i/K_R^i$,
$\frac{[T_NA_i]}{[R_N]}=L\frac{N!}{i!(N-i)!}[A]^i/K_T^i$,
$i=1,2,\cdots,N$. The fractional saturation

\begin{equation}\label{Monod}
\phi
=\frac{(1+[A]/K_R)^{N-1}[A]/K_R+L(1+[A]/K_T)^{N-1}[A]/K_T}{(1+[A]/K_R)^N+L(1+[A]/K_T)^N},
\end{equation}
which can be also rearranged as the Adair equation, where the
dissociation constants
$K_i=\frac{\frac{1}{K_R^{i-1}}+\frac{L}{K_T^{i-1}}}{\frac{1}{K_R^{i}}+\frac{L}{K_T^{i}}}$,
$i=1,2,\cdots,N$.

Similar to the case of two site, one can derive $K_i\geq K_{i+1}$,
which also implies positive cooperativity.

When $K_R\neq K_T$, the steepness of the curve passes through a
maximum when $L^2=\frac{K_T^N}{K_R^N}$ \mycite{CB}. The
half-saturation concentration $K_{0.5}=\sqrt{K_RK_T}$ and the Hill
coefficient $n_H=2\frac{d\log\phi }{d\log[A]}|_{\phi
=\frac{1}{2}}=N-\frac{4(N-1)\sqrt{\frac{K_T}{K_R}}}{(1+\sqrt{\frac{K_T}{K_R}})^2}$.

\subsection{Sequential model}

Koshland, Nemethy and Filmer  \mycite{KNF} showed a more orthodox
application of induced fit theory  \mycite{Kosh58,Kosh59a,Kosh59b},
known as the sequential model. They also postulated the existence of
two conformations, but one of them is induced by ligand binding.

\subsubsection{Dimer}

Sequential model of a two-site protein is illustrated in Fig.
\ref{fig7}, recapitulated also from \mycite{CB}.

Basic parameters: $K_t$ is the notional equilibrium constant of the
conformation change $T\rightarrow R$ ($K_t=[T]/[R]\gg 1$), and $K_A$
is the dissociation constant of the conformation $R$ bounded with a
molecule of the substrate $A$.  Moreover, in order to consider the
interface across change, we should introduce the parameters
$K_{R:T}$ and $K_{R:R}$, representing the notional equilibrium
constants for the interface of the two sites changing from $T:T$ to
$R:T$ and $R:R$ respectively.

Hence, $[TRA]=\frac{2[T_2][A]K_{R:T}}{K_tK_A}$ and
$[R_2A_2]=\frac{[TRA][A]K_{R:R}}{2K_tK_AK_{R:T}}=\frac{[T_2][A]^2K_{R:R}}{K_t^2K_A^2}$,
which give rise to the fractional saturation

$$\phi =\frac{[TRA]+2[R_2A_2]}{2([T_2]+[TRA]+[R_2A_2])}=\frac{\frac{[A]K_{R:T}}{K_tK_A}+\frac{[A]^2K_{R:R}}{K_t^2K_A^2}}{1+\frac{2[A]K_{R:T}}{K_tK_A}+\frac{[A]^2K_{R:R}}{K_t^2K_A^2}}.$$

Let $c^2=\frac{K_{R:T}^2}{K_{R:R}}$ and
$\bar{K}=\frac{K_tK_A}{K_{R:R}^{\frac{1}{2}}}$, then
$$\phi =\frac{c[A]/\bar{K}+[A]^2/\bar{K}^2}{1+2c[A]/\bar{K}+[A]^2/\bar{K}^2},$$
which is an Adair equation with the dissociation constants
$K_1=\bar{K}/c$ and $K_2=c\bar{K}$. Consequently, $c>1$ implies the
negative cooperativity, while $c<1$ implies the positive
cooperativity.

\subsubsection{Quaternary structure}

Basic parameter: $K$ is the equilibrium constant of single site
bound with a substrate molecule $A$, and $y$ represents the
interaction energy (similar to the famous work of Pauling
 \mycite{Pau35}).

Hence the fractional saturation
$$\phi =\frac{4K[A]+2(4K^2[A]^2y+2K^2[A]^2)+12K^3[A]^3y^2+4K^4[A]^4y^4}{4(1+4K[A]+4K^2[A]^2y+2K^2[A]^2+4K^3[A]^3y^2+K^4[A]^4y^4)},$$
which can also be expressed as the Adair equation with the Adair
constants $K_1=\frac{1}{K}$, $K_2=\frac{3}{(2y+1)K}$,
$K_3=\frac{2y+1}{3y^2K}$ and $K_4=\frac{1}{Ky^2}$. Hence, if $y>1$,
there is a positive cooperativity, and if $y<1$, there is a negative
cooperativity.

It is just the example used by Hong Qian  \mycite{QH2007} in order
to explain the relationship between temporal and allosteric
cooperativity phenomena. But the analysis there is somewhat vague
and incomplete.

\section{Discussion}

Nowadays, an era of quantifying the signaling processes in terms of
physiochemical principles is emerging  \mycite{Kosh98,HHLM}.
Quantitative understanding and mathematical modeling of biological
systems presents a significant challenge as well as an unique
opportunity for scientists of diverse disciplines.

During the theoretical development of signal transduction network,
sensitivity plays an indispensable role, and the mechanism of high
sensitivity, for instance the zero-order ultrasensitivity
\mycite{GK}, may be needed for the adaptive sensory systems, in
which one pathway must be turned on and another pathway turned off.

Although the sharp activation in PdPC switches have always been
compared to allosteric cooperative transitions \mycite{KGS}, it has
never been made very clear what the essential similarities and
differences between them are. This significant question could date
back to Fischer and Krebs \mycite{Fisch55,Fisch71}, who discovered
protein phosphorylation as a regulatory mechanism for enzyme
activity and won the Nobel Prize in 1992.

While the requirements for both nonlinearity and nonequilibrium are
intuitively obvious \mycite{NP,Mu}, quantitative aspects of such a
system have never been studied until Qian's work
\mycite{QH2003,QH2007}, which answered one aspect of this basic
question. He suggested that the essential difference between the
allosteric mechanism and the hydrolysis cycle is that the former
does not expend energy: ``The costs of the two types of regulations
are quite different. One requires a significant amount of regulator
biosynthesis in advance. The other requires only a small amount of
regulators for the hydrolysis reaction, but it consumes energy
during the regulation.''\mycite{QH2007}

The thermodynamic analysis for the phosphorylation-dephosphorylation
cycle (PdPC) is provided (See Section 2 and Appendix) to confirm the
conclusion that $\gamma$ is the unique control parameter for the
nonequilibrium steady state. Then in Section 3, it is shown that the
key result in Ref. \mycite{QH2003} also holds in the stochastic
model, which implies that the PdPC switch is a phenomenon only
exhibited in nonequilibrium steady states.

Our quantitative analysis provided a clear mechanistic origin for
the high cooperativity in the zero-order ultrasensitivity. A reduced
chemical master equation (Fig.  \ref{fig_Reduced_stoch}) indicates
that the mechanism of temporal cooperativity is parallel in
mathematical form to, but fundamentally different in biochemical
nature from, the allosteric cooperativity of multi-subunits protein
systems, where the dissociation constants play the key role.

Nevertheless, the degree of allosteric cooperativity is restricted
by the total number of sites in a single enzyme molecule which can
not be freely regulated, while temporal cooperativity is only
restricted by the total molecule number of the target protein which
can be regulated in a wide range and gives rise to the
ultrasensitivity phenomenon. That is just why the organisms find it
advantageous to develop the mechanism of covalent modification via
phosphorylation and $ATP$ hydrolysis to control the biological
activity of proteins rather than the mechanism of allosteric
transitions.

Therefore, the improving of the total number of molecules of target
protein can not increase the degree of allosteric cooperativity,
while it can obviously increase the degree of temporal cooperativity
, indicated by the increasing gradients of the fractional saturation
function $\langle \phi \rangle$ (Fig. \ref{Fig_ultra_switch_fig3})
and the decreasing dissociation constants $\{K_j,~j=1,2,\cdots,N\}$
(Fig. \ref{Fig_ultra_switch_coop})!

On the other hand, the present research also emphasizes that
nonlinearity of the forward and backward fluxes is another
requirement for sharp transitions with ultrasensitivity. Moreover,
we express the nonlinearity by the varying of dissociation
constants, which exhibits the essential difference between the
simple and ultrasensitive PdPC switches (See Fig.
\ref{Fig_simple_switch_coop} and Fig.  \ref{Fig_ultra_switch_coop}).

Finally, it is often thought that the noise added to the biological
models only provides moderate refinements to the behaviors otherwise
predicted by the classical deterministic system description, while
in the present paper, it is quite clear that the main result, namely
the mathematical equivalence between temporal and allosteric
cooperativity can only be explicitly expressed by the chemical
master equation model (See Fig.  \ref{fig_Reduced_stoch}), where
nonequilibrium is hidden in the parameter $\gamma\neq 1$.

The concept of temporal cooperativity in terms of the random-walk
model is not limited to PdPC and kinetically isomorphic GTPases, but
also applies to many other signaling processes
\mycite{QH_BJREV2008}.

\section*{Acknowledgment}

The authors would like to thank Prof. Hong Qian in University of
Washington and Prof. Minping Qian, Prof. Xufeng Liu in Peking
University for helpful discussions. After completion of the present
work, we have received a preprint (Ref. \mycite{QH_BJREV2008}) on a
similar problem, but our focuses are quite different. This work is
partly supported by the NSFC (Nos. 10701004, 10531070 and 10625101)
and 973 Program 2006CB805900.

\small
\section{Appendix}

\subsection{Complete mathematical models and nonequilibrium steady states}

\subsubsection{Deterministic model: mass action law}

Biologists usually build the deterministic model of biochemical
systems from the macroscopic view. Based on the mass action law, the
forward and backward fluxes of chemical reaction $A1$ are
$J_1=a_1[W][E_1]$ and $J_{-1}=d_1[WE_1]$ respectively; similarly,
the forward and backward fluxes of chemical reactions $A2$, $A3$ and
$A4$ are $J_2=k_1[WE_1]$, $J_{-2}=q_1[W^*][E_1]$,
$J_3=a_2[W^*][E_2]$, $J_{-3}=d_2[W^*E_2]$, $J_4=k_2[W^*E_2]$ and
$J_{-4}=q_2[W][E_2]$ respectively.

We can choose $[W^*]$, $[E_1]$ and $[E_2]$ as independent variables
according to the three restrictions $W_T=[W]+[WE_1]+[W^*E_2]+[W^*]$,
$E_{1T}=[E_1]+[WE_1]$ and $E_{2T}=[E_2]+[W^*E_2]$, where $W_T$,
$E_{1T}$ and $E_{2T}$ are constants representing the total
concentrations of target protein, kinase and phosphatase
respectively. Then the deterministic equations are

\begin{eqnarray}\label{Complete-deter-model}
\frac{d[W^*]}{dt}&=&J_2-J_{-2}+J_{-3}-J_3;\nonumber\\
\frac{d[E_1]}{dt}&=&J_{-1}-J_1+J_2-J_{-2};\nonumber\\
\frac{d[E_2]}{dt}&=&J_{-3}-J_3+J_4-J_{-4}.
\end{eqnarray}

In the steady state, the right side of (\ref{Complete-deter-model})
is set to be zero, which leads to the important definition of the
net flux $J\stackrel{def}{=}J_i-J_{-i},~i=1,2,3,4$. Based on the
relation
$\gamma\stackrel{def}{=}\frac{a_1k_1a_2k_2}{d_1q_1d_2q_2}=\frac{J_1J_2J_3J_4}{J_{-1}J_{-2}J_{-3}J_{-4}}$,
we will know that $J>0$ is equivalent to the energy parameter
$\gamma>1$; and $J<0$ is equivalent to $\gamma<1$. Moreover, the
entropy production which is a key concept in nonequilibrium
thermodynamics can be expressed as
$$ep=flux\times potential=J\dot\log\gamma.$$

Obviously, $ep=0$ if and only if $\gamma=1$, which means chemical
equilibrium state according to the thermodynamic analysis in Section
\ref{Sec-noneq}.

In addition, it is necessary to note that we have a nonlinear
system, where the well-known King-Altman method  \mycite{CB} fails.

\subsubsection{Chemical master equation of the complete model}

A deterministic model, however, only describes the averaged behavior
of a system of large populations, and can not capture the temporal
fluctuations of a small biological system with either extrinsic or
intrinsic noise. Hence stochastic models with chemical master
equations (CME) based on biochemical reaction stoichiometry,
molecular numbers, and kinetic rate constants are worth being
applied.

Denote the volume as $V$, which is a fixed parameter of the system.
And let $N_T=W_TV$, $N_{1T}=E_{1T}V$ and $N_{2T}=E_{2T}V$, recalling
$W_T=[W]+[WE_1]+[W^*E_2]+[W^*]$, $E_{1T}=[E_1]+[WE_1]$ and
$E_{2T}=[E_2]+[W^*E_2]$ are constants representing the total
concentrations of target protein, kinase and phosphatase
respectively. So we can still choose the molecule numbers of species
$W^*$, $E_1$ and $E_2$ as three independent variables. Let
$P(i,j,k;t)$ be the probability of the event that the molecule
numbers of species $W^*$, $E_1$ and $E_2$ at time $t$ are $i,j$ and
$k$ respectively, which satisfies the {\bf chemical master equation}

\begin{eqnarray}\label{Complete-stoch-model}
&&\frac{dP(i,j,k;t)}{dt}\nonumber\\
&=&\frac{a_1}{V}(N_T-N_{1T}-N_{2T}-i+j+k+1)(j+1)P(i,j+1,k;t)+d_1(N_{1T}-j+1)P(i,j-1,k;t)\nonumber\\
&+&k_1(N_{1T}-j+1)P(i-1,j-1,k;t)+\frac{q_1}{V}(i+1)(j+1)P(i+1,j+1,k;t)\nonumber\\
&+&\frac{a_2}{V}(i+1)(k+1)P(i+1,j,k+1;t)+d_2(N_{2T}-k+1)P(i-1,j,k-1;t)\nonumber\\
&+&k_2(N_{2T}-k+1)P(i,j,k-1;t)+\frac{q_2}{V}(N_T-N_{1T}-N_{2T}-i+j+k+1)(k+1)P(i,j,k+1;t)\nonumber\\
&-&[\frac{a_1}{V}(N_T-N_{1T}-N_{2T}-i+j+k)j+d_1(N_{1T}-j)+k_1(N_{1T}-j)+\frac{q_1}{V}ij\nonumber\\
&&+\frac{a_2}{V}ik+d_2(N_{2T}-k)+k_2(N_{2T}-k)+\frac{q_2}{V}(N_T-N_{1T}-N_{2T}-i+j+k)k]P(i,j,k;t).
\end{eqnarray}

It is necessary to explain the discrete population coefficients in
the above equation. When the system is in the state $(i,j,k)$, the
molecular numbers of $WE_1$, $W^*E_2$ and $W$ are $N_{1T}-j$,
$N_{2T}-k$ and $N_T-i-(N_{1T}-j)-(N_{2T}-k)$ respectively. Moreover,
the parameters $\frac{a_1}{V}$, $\frac{a_2}{V}$, $\frac{q_1}{V}$ and
$\frac{q_2}{V}$ are called  ``stochastic rate constants''
\mycite{WD}, and their relationships with the original rate
constants $a_1$, $a_2$, $q_1$ and $q_2$ have been developed in Ref.
\mycite{Kur}. For instance, the quantity $J_1=a_1[W][E_1]$ is in the
unit of concentration, hence the stochastic rate $J_1\times
V=a_1V\times\frac{N_T-N_{1T}-N_{2T}-i+j+k}{V}\times\frac{j}{V}=\frac{a_1}{V}(N_T-N_{1T}-N_{2T}-i+j+k)j$
should be in the unit of molecular numbers when we build chemical
master equations.

This is a continuous-time jumping process on the three-dimensional
cube $N_T\times N_{1T}\times N_{2T}$. The state $(i,j,k)$ can only
jump to the adjacent states $(i,j+1,k)$, $(i,j-1,k)$, $(i-1,j-1,k)$,
$(i+1,j+1,k)$, $(i+1,j,k+1)$, $(i-1,j,k-1)$, $(i,j,k-1)$ and
$(i,j,k+1)$.

In probability theory, such a random-walk model is called the
three-dimensional birth-and-death process, which is a special Markov
chain. Generally speaking, $\xi$ and $\eta$ represent the states and
$q_{\xi\eta}$ is the transition density along the passage
$\xi\rightarrow\eta$. The equation (\ref{Complete-stoch-model}) is
just the Kolmogorov forward equation (also called the Fokker-Planck
equation) of the continuous-time Markov chain with transition
density matrix $Q=(q_{\xi\eta})$

\begin{equation}
\frac{dP(\xi,t)}{dt}=P(\xi,t) Q,
\end{equation}
where $\xi=(\xi^1,\xi^2,\xi^3)$ represents the state in which the
molecule numbers of $W^*$, $E_1$ and $E_2$ are $\xi^1$, $\xi^2$ and
$\xi^3$ respectively, and
$$q_{\xi\eta}=\left\{\begin{array}{ll}\frac{a_1}{V}(N_T-N_{1T}-N_{2T}-i+j+k)j&\xi=(i,j,k),\eta=(i,j+1,k),\\d_1(N_{1T}-j)&\xi=(i,j,k),\eta=(i,j-1,k),\\k_1(N_{1T}-j)&\xi=(i,j,k),\eta=(i-1,j-1,k),
\\\frac{q_1}{V}ij&\xi=(i,j,k),\eta=(i+1,j+1,k),\\\frac{a_2}{V}ik&\xi=(i,j,k),\eta=(i+1,j,k+1),\\d_2(N_{2T}-k)&\xi=(i,j,k),\eta=(i-1,j,k-1),\\k_2(N_{2T}-k)&\xi=(i,j,k),\eta=(i,j,k-1),
\\\frac{q_2}{V}(N_T-N_{1T}-N_{2T}-i+j+k)k&\xi=(i,j,k),\eta=(i,j,k+1),\\-\sum_{\zeta\neq \xi}q_{\xi\zeta}&\xi=\eta=(i,j,k),\\0&else\end{array}\right.$$

\subsubsection{Rigorous thermodynamic analysis}\label{Sec-noneq}

1. From the perspective of the deterministic(macroscopic) model, the
system is in equilibrium state, if and only if the forward and
backward fluxes of each chemical reaction are equal, i.e.
$J_1=J_{-1}$, $J_2=J_{-2}$, $J_3=J_{-3}$ and $J_4=J_{-4}$. Hence
$\gamma=1$ is necessary for the equilibrium state.

For the sufficiency, we have to show that if $\gamma=1$, there
exists an unique reasonable solution under the equilibrium
conditions $J_1=J_{-1}$, $J_2=J_{-2}$, $J_3=J_{-3}$ and
$J_4=J_{-4}$.

Since $J_1=J_{-1}$ implies $[WE_1]=\frac{a_1[W]E_{1T}}{d_1+a_1[W]}$,
$J_3=J_{-3}$ implies $[W^*E_2]=\frac{a_2[W^*]E_{2T}}{d_2+a_2[W^*]}$
and $J_1J_2=J_{-1}J_{-2}$ implies $[W^*]=\frac{a_1k_1}{d_1q_1}[W]$;
then, the equality $W_T=[W]+[WE_1]+[W^*E_2]+[W^*]$ becomes
$$W_T=[W]+\frac{a_1[W]E_{1T}}{d_1+a_1[W]}+\frac{a_2\frac{a_1k_1}{d_1q_1}[W]E_{2T}}{d_2+a_2\frac{a_1k_1}{d_1q_1}[W]}+\frac{a_1k_1}{d_1q_1}[W].$$

The right side is an increasing function of $[W]$, and it equals
zero when $[W]=0$ and is larger than $W_T$ when $[W]=W_T$. Hence the
above equation has an unique reasonable solution between $0$ and
$W_T$.

Finally, it could be rigorously proved that the ordinary
differential equations (\ref{Complete-deter-model}) only have an
unique fixed point (See Section \ref{Appen-unique}), which finishes
our proof for sufficiency.

2. From the perspective of the stochastic (mesoscopic) model, we
should appeal to the chemical master equation
(\ref{Complete-stoch-model}). In the mathematical theory of
nonequilibrium steady states  \mycite{Sc, JQQb}, there is a famous
condition named ``Kolmogorov's cyclic condition''(See Section
\ref{rev-cir-Kol}), which is equivalent to the
reversibility(equilibrium) of the specific Markov chain. The
priority of this condition is that one can directly write down the
condition for reversibility without deriving the steady states
first. Although there are many many cycles in the Markov chain model
(\ref{Complete-stoch-model}), every large cycle can be decomposed
into several basic four-state cycles
$$\xi_1=(i,j,k)\rightarrow \xi_2=(i,j-1,k)\rightarrow \xi_3=(i+1,j,k)\rightarrow \xi_4=(i,j,k-1)\rightarrow \xi_1=(i,j,k),$$
which just accords to the kinetic phosphorylation-dephosphorylation
cycle.

In this case, the necessary and sufficient condition for the steady
state being in equilibrium, i.e. the Kolmogorov cyclic condition, is
expressed as
$q_{\xi_1\xi_2}q_{\xi_2\xi_3}q_{\xi_3\xi_4}q_{\xi_4\xi_1}=q_{\xi_1\xi_4}q_{\xi_4\xi_3}q_{\xi_3\xi_2}q_{\xi_2\xi_1}$.
From (\ref{Complete-stoch-model}), this is just

\begin{eqnarray}
&&\frac{a_1}{V}(N_T-N_{1T}-N_{2T}-i+j+k)j\times k_1(N_{1T}-j+1)
\times \frac{a_2}{V}(i+1)k\times k_2(N_{2T}-k+1)\nonumber\\
&=&\frac{q_2}{V}(N_T-N_{1T}-N_{2T}-i+j+k)k\times d_2(N_{2T}-k+1)
\times \frac{q_1}{V}(i+1)j\times d_1(N_{1T}-j+1).\nonumber
\end{eqnarray}
Hence one can derive that
$\gamma\stackrel{\triangle}{=}\frac{a_1k_1a_2k_2}{d_1q_1d_2q_2}=1$.

Namely, $\gamma\neq 1$ is equivalent to the fact that this system is
in a nonequilibrium steady state.

\subsection{The complete and reduced models share the same steady state}\label{Appen-samess}

The rationality of the reduced model (\ref{Reduced_gen_model}) is
based on the fact that its steady state satisfying
$f_1([W]^{ss})=f_2([W^*]^{ss})$ is the same as that of the complete
model (\ref{Complete-deter-model}), under the restriction
$W_T=[W]^{ss}+[W^*]^{ss}$!

The steady state of the complete model (\ref{Complete-deter-model})
satisfies that
\begin{equation}\label{switch_ss1}
k_1[WE_1]-q_1[W^*]^{ss}[E_1]+d_2[W^*E_2]-a_2[W^*]^{ss}[E_2]=0,
\end{equation}

\begin{equation}\label{switch_ss2}
d_1[WE_1]-a_1[W]^{ss}[E_1]+k_1[WE_1]-q_1[W^*]^{ss}[E_1]=0,
\end{equation}

\begin{equation}\label{switch_ss3}
d_2[W^*E_2]-a_2[W^*]^{ss}[E_2]+k_2[W^*E_2]-q_2[W]^{ss}[E_2]=0.
\end{equation}

From (\ref{switch_ss2}),
$$[E_1]=\frac{(d_1+k_1)E_{1T}}{d_1+k_1+a_1[W]^{ss}+q_1[W^*]^{ss}},~[WE_1]=\frac{(a_1[W]^{ss}+q_1[W^*]^{ss})E_{1T}}{d_1+k_1+a_1[W]^{ss}+q_1[W^*]^{ss}};$$

and from (\ref{switch_ss3}),
$$[E_2]=\frac{(d_2+k_2)E_{2T}}{d_2+k_2+q_2[W]^{ss}+a_2[W^*]^{ss}},~[W^*E_2]=\frac{(q_2[W]^{ss}+a_2[W^*]^{ss})E_{2T}}{d_2+k_2+q_2[W]^{ss}+a_2[W^*]^{ss}};$$
which combined with (\ref{switch_ss1}) gives

\begin{eqnarray}
&&k_1\frac{(a_1[W]^{ss}+q_1[W^*]^{ss})E_{1T}}{d_1+k_1+a_1[W]^{ss}+q_1[W^*]^{ss}}-q_1[W^*]^{ss}\frac{(d_1+k_1)E_{1T}}{d_1+k_1+a_1[W]^{ss}+q_1[W^*]^{ss}}\nonumber\\
&=&a_2[W^*]^{ss}\frac{(d_2+k_2)E_{2T}}{d_2+k_2+q_2[W]^{ss}+a_2[W^*]^{ss}}-d_2\frac{(q_2[W]^{ss}+a_2[W^*]^{ss})E_{2T}}{d_2+k_2+q_2[W]^{ss}+a_2[W^*]^{ss}},\nonumber
\end{eqnarray}
which is just the equation $f_1([W]^{ss})=f_2([W^*]^{ss})$.

\subsection{Derivation of the steady distribution in the reduced stochastic model}\label{Appen-ss}

Let the right side of (\ref{Reduced-stoch-model}) equals zero, which
gives
\begin{eqnarray}
&&f_2(1/V)VP^{ss}(N-1,1)-f_1(N/V)VP^{ss}(N,0)=0;\nonumber\\
&&f_1((N+1-i)/V)VP^{ss}(N+1-i,i-1)+f_2((i+1)/V)VP^{ss}(N-1-i,i+1)\nonumber\\
&&-[f_1((N-i)/V)+f_2(i/V)]VP^{ss}(N-i,i)=0,~i=1,2,\cdots,N-1;\nonumber\\
&&f_1(1/V)VP^{ss}(1,N-1)-f_2(N/V)VP^{ss}(0,N)=0.\nonumber
\end{eqnarray}

So
\begin{eqnarray}
&&f_1(N/V)VP^{ss}(N,0)=f_2(1/V)VP^{ss}(N-1,1);\nonumber\\
&&f_1((N+1-i)/V)VP^{ss}(N+1-i,i-1)-f_2(i/V)VP^{ss}(N-i,i)\nonumber\\
&&=f_1((N-i)/V)VP^{ss}(N-i,i)-f_2((i+1)/V)VP^{ss}(N-1-i,i+1),~i=1,2,\cdots,N-1;\nonumber\\
&&f_1(1/V)VP^{ss}(1,N-1)=f_2(N/V)VP^{ss}(0,N).\nonumber
\end{eqnarray}

Then applying the iteration technique, we have
$$f_1((N+1-i)/V)VP^{ss}(N+1-i,i-1)=f_2(i/V)VP^{ss}(N-i,i),~i=1,2,\cdots,N,$$
which means in the steady state, the ratio of the probabilities of
the states $(N-i,i)$ and $(N,0)$ is
$\Pi_{j=1}^i\left[\frac{f_1((N+1-j)/V)V}{f_2(j/V)V}\right]=\Pi_{j=1}^i\left[\frac{f_1((N+1-j)/V)}{f_2(j/V)}\right]$.

Consequently, the steady distribution of $(N-i,i)$ is
\begin{eqnarray}
P^{ss}(N-i,i)=\frac{\Pi_{j=1}^i\frac{f_1((N+1-j)/V)}{f_2(j/V)}}{1+\sum_{i=1}^N\Pi_{j=1}^i\frac{f_1((N+1-j)/V)}{f_2(j/V)}}.\nonumber
\end{eqnarray}

\subsection{Existence of the unique reasonable solution in the deterministic model of the PdPC
switch}\label{Appen-unique}

Based on the analysis in Section \ref{Appen-samess}, we have already
known the steady solutions of the complete model
(\ref{Complete-deter-model}) and reduced simple model
(\ref{Reduced_gen_model}) are the same, i.e. both satisfying
$$f_1([W]^{ss})=f_2([W^*]^{ss}).$$

On the other hand, according to the analysis in Section
\ref{Sec_reduced-deter-model}, we also have known that under the
assumption $W_T\gg E_{1T}+E_{2T}$ (i.e. $W_T=[W]+[W^*]$), $\phi
=\frac{[W^*]^{ss}}{W_T}$ satisfies

$$\theta=\frac{\mu\gamma[\mu-(\mu+1)\phi ](\phi -\frac{K_1^*(W_T+K_1)}{(K_1^*-K_1)W_T})K_2K_2^*(K_1^*-K_1)}{[\mu\gamma-(\mu\gamma+1)\phi ](\phi +\frac{K_2^*(W_T+K_2)}{(K_2-K_2^*)W_T})K_1K_1^*(K_2-K_2^*)}.$$

Define
\begin{eqnarray}
g(\phi )&=&\mu\gamma[\mu-(\mu+1)\phi ](\phi -\frac{K_1^*(W_T+K_1)}{(K_1^*-K_1)W_T})K_2K_2^*(K_1^*-K_1)\nonumber\\
&&-\theta [\mu\gamma-(\mu\gamma+1)\phi ](\phi
+\frac{K_2^*(W_T+K_2)}{(K_2-K_2^*)W_T})K_1K_1^*(K_2-K_2^*),\nonumber
\end{eqnarray}
which is a quadratic equation. Hence we only need to prove $g(0)<0$
and $g(1)>0$.

\begin{eqnarray}
g(0)&=&\mu\gamma\mu(-\frac{K_1^*(W_T+K_1)}{(K_1^*-K_1)W_T})K_2K_2^*(K_1^*-K_1)\nonumber\\
&&-\theta
\mu\gamma\frac{K_2^*(W_T+K_2)}{(K_2-K_2^*)W_T}K_1K_1^*(K_2-K_2^*)<0\nonumber
\end{eqnarray}
is obvious.

And
\begin{eqnarray}
g(1)&=&\mu\gamma(-1)(1-\frac{K_1^*(W_T+K_1)}{(K_1^*-K_1)W_T})K_2K_2^*(K_1^*-K_1)\nonumber\\
&&-\theta
(-1)(1+\frac{K_2^*(W_T+K_2)}{(K_2-K_2^*)W_T})K_1K_1^*(K_2-K_2^*)\nonumber
\end{eqnarray}
is also obvious, because $\frac{K_1^*(W_T+K_1)}{(K_1^*-K_1)W_T}>1$.

Therefore, there is only one solution of $g(\phi )=0$ in the
interval $[0,1]$.

\subsection{Kolmogorov cyclic condition}\label{rev-cir-Kol}

This subsection is recapitulated from \mycite{JQQb}.

Suppose that $X$ is an irreducible and positive-recurrent stationary
Markov chain with the countable state space $S$, the transition
density matrix $Q=(q_{ij})_{i,j\in S}$ and the invariant probability
distribution $\Pi=(\pi_i)_{i\in S}$, then the following statements
are
equivalent: \\
(i) The Markov chain $X$ is reversible (equilibrium). \\
(ii) The Markov chain $X$ is in detailed balance, that is,
 $$\pi_i q_{ij}=\pi_j q_{ji},\forall i,j\in S.$$
(iii) The transition probability of $X$ satisfies the {\bf
Kolmogorov cyclic condition}:
 $$q_{i_1i_2}q_{i_2i_3}\cdots q_{i_{s-1}i_s}q_{i_si_1}
   =q_{i_1i_s}q_{i_si_{s-1}}\cdots q_{i_3i_2}q_{i_2i_1},$$
for any directed cycle $c=(i_1,\cdots,i_s)$. \\

\newpage

\section*{List of Figure Captions}

\renewcommand{\baselinestretch}{2}


Fig 1: {\sl The reduced model of PdPC switch.}

Fig 2: {\sl The illustrated chemical master equation of the reduced
model of the PdPC switch. The two dimensional vector $(N-i,i)$
represents the random state that the molecule number of the species
$W$ is $(N-i)$ and the molecule number of the species $W^*$ is $i$.}

\begin{figure}[h]
\centerline{\includegraphics[width=4in,height=3in]{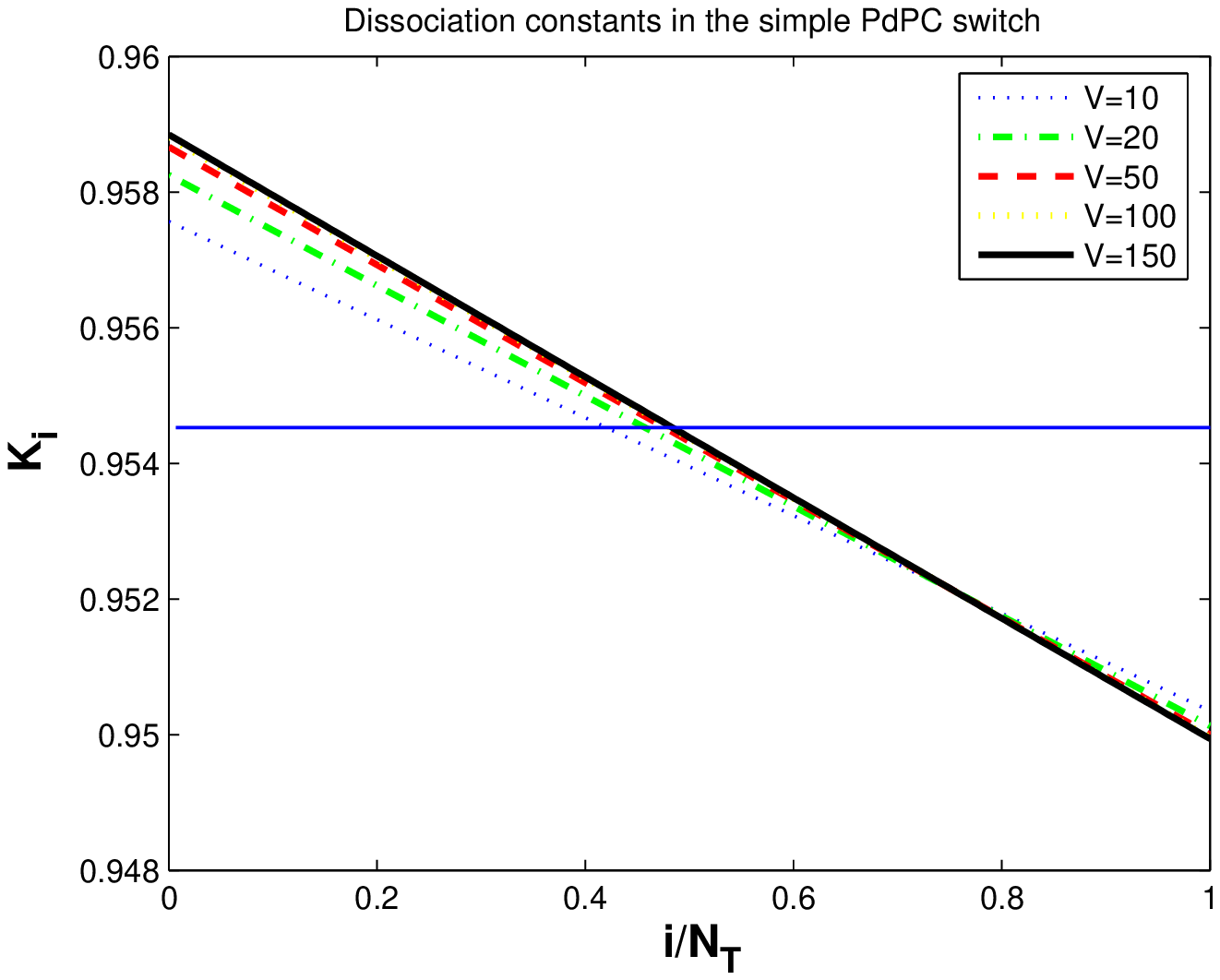}}
\caption[Fig_simple_switch_coop]{} \label{Fig_simple_switch_coop}
\end{figure}

\begin{figure}[h]
\centerline{\includegraphics[width=4in,height=3in]{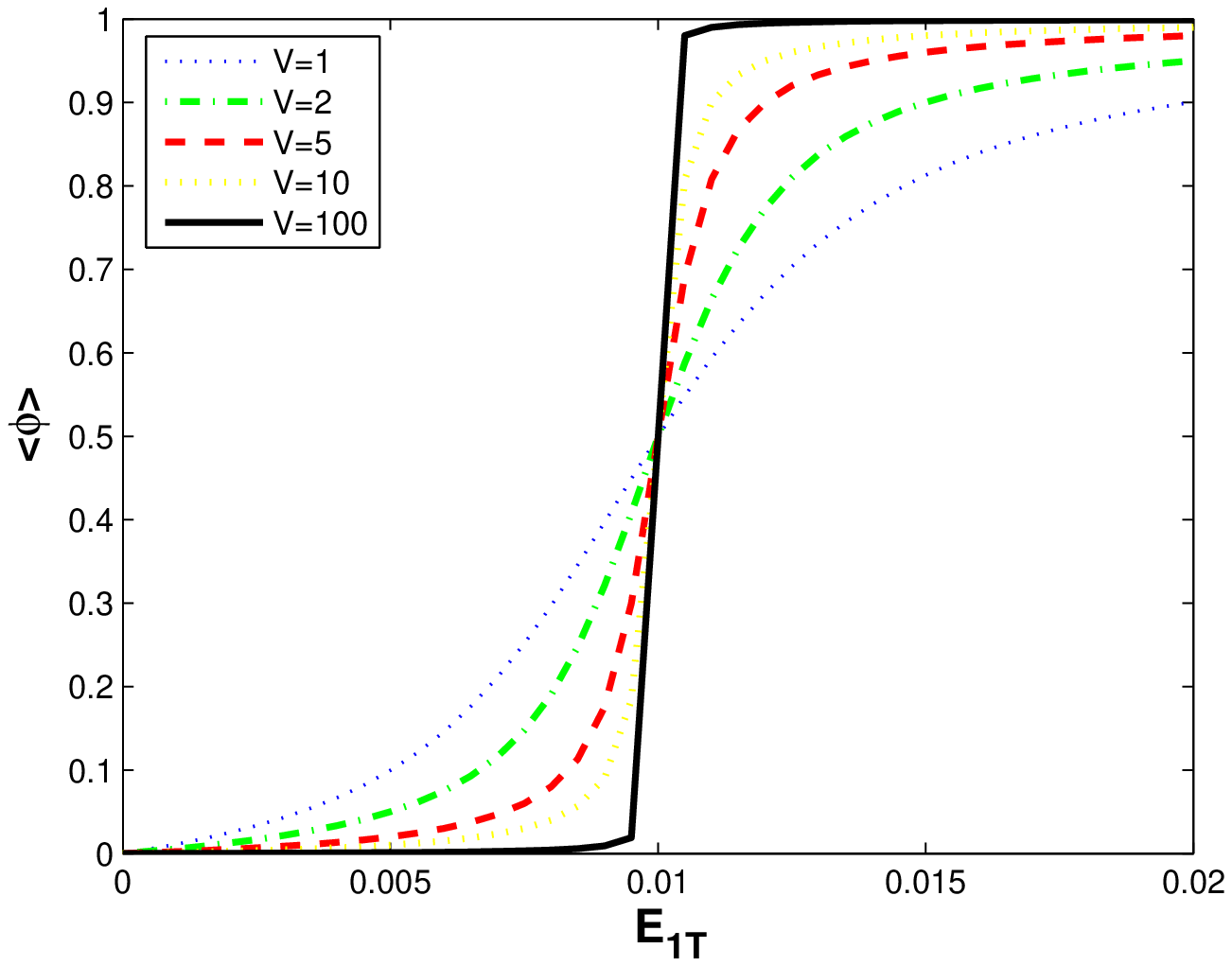}}
\caption[Fig_ultra_switch_fig2]{} \label{Fig_ultra_switch_fig2}
\end{figure}

\begin{figure}[h]
\centerline{\includegraphics[width=4in,height=3in]{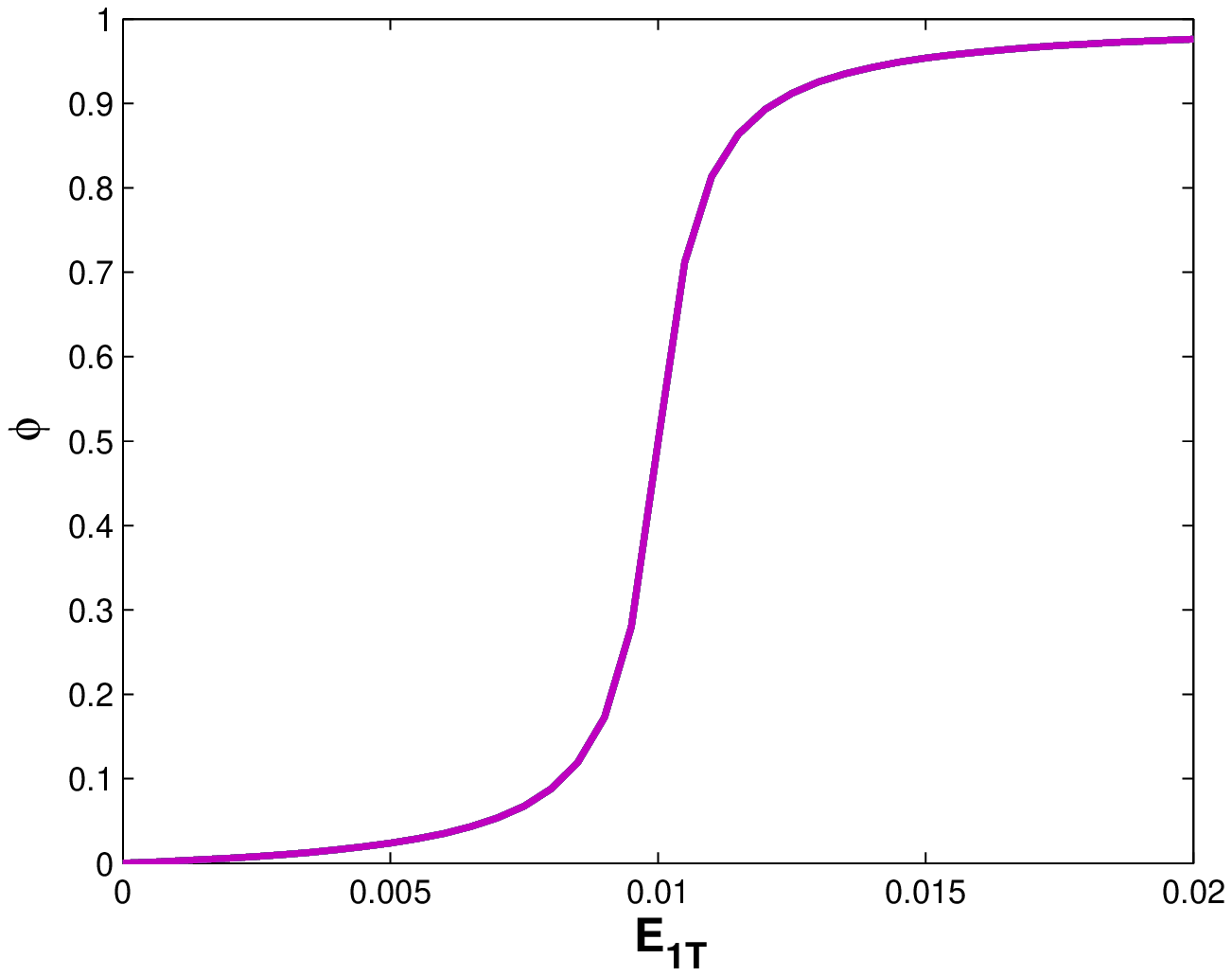}}
\caption[Fig_ultra_switch_fig3_1]{} \label{Fig_ultra_switch_fig3_1}
\end{figure}

\begin{figure}[h]
\centerline{\includegraphics[width=4in,height=3in]{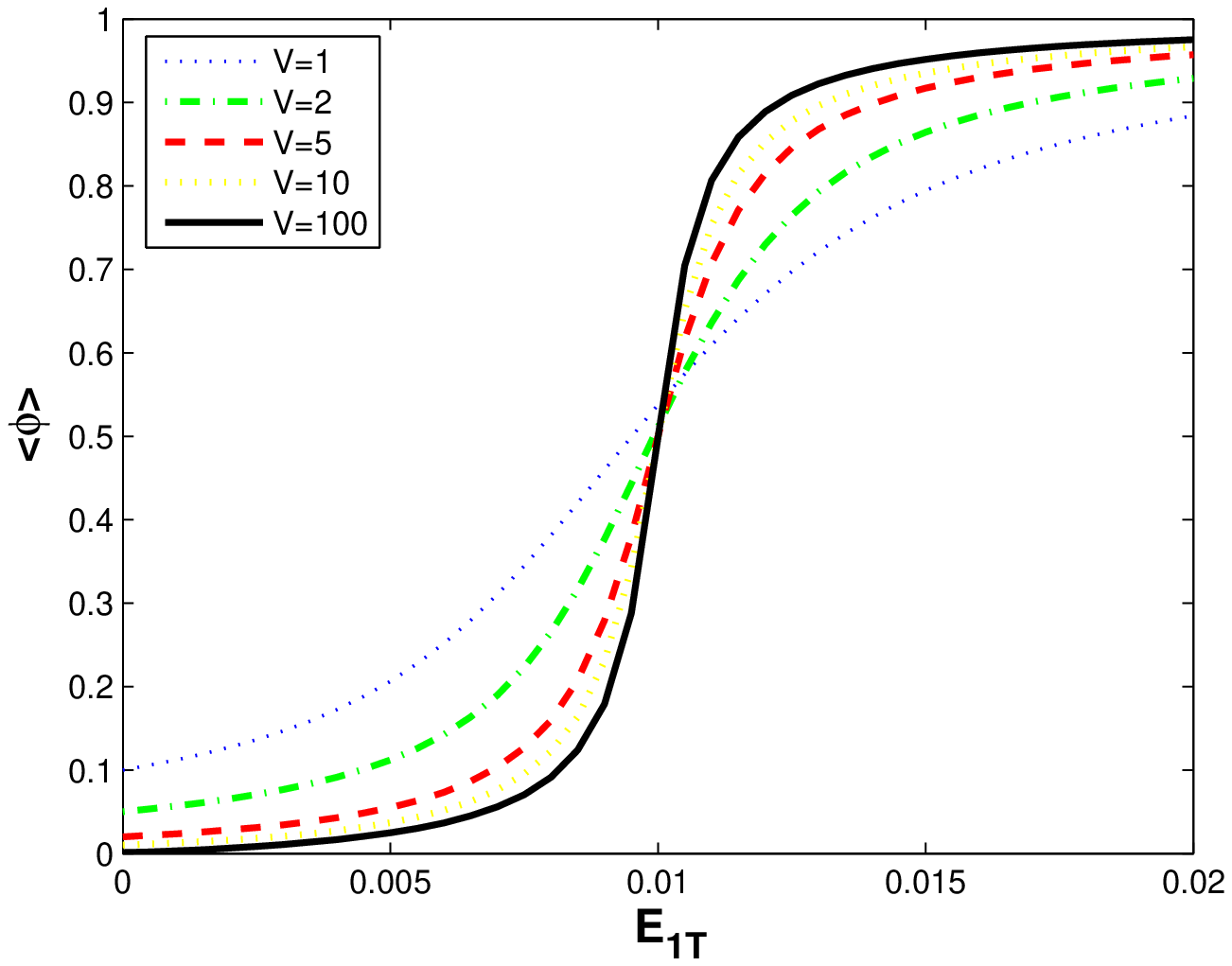}}
\caption[Fig_ultra_switch_fig3]{} \label{Fig_ultra_switch_fig3}
\end{figure}

Fig 3: {\sl The curve of $\phi $ with respect to $E_{1T}$ in the
deterministic model of the simple PdPC switch without the
first-order linear approximation, where the parameters are the same
as that in Fig. \ref{Fig_simple_switch_coop}.}

Fig 4: {\sl The curve of $\langle \phi \rangle $ with respect to
$E_{1T}$ in the stochastic model of the simple PdPC switch without
the first-order linear approximation at different volumes, where the
parameters are the same as that in Fig.
\ref{Fig_simple_switch_coop}.}

Fig 5: {\sl The dissociation constants in the simple PdPC switch
with different volumes, where $a_1=0.01; d_1=1; k_1=1; q_1=0.0001;
E_{1T}=0.01; a_2=0.01; d_2=1; k_2=1; q_2=0.0001; E_{2T}=0.01;
W_T=1$, and $\alpha=(V_1/K_1+V_2/K_2)/(V_1/K_1+V_2/K_2)=1$. The
volume $V$ takes different values as 10, 20, 50, 100 and 150, namely
the total molecule number $N=W_TV$ takes values 10, 20, 50, 100 and
150 respectively. The horizontal line represents the quantity
$1/\alpha$, which equals all the dissociation constants under the
first-order assumption.}

\begin{figure}[h]
\centerline{\includegraphics[width=4in,height=3in]{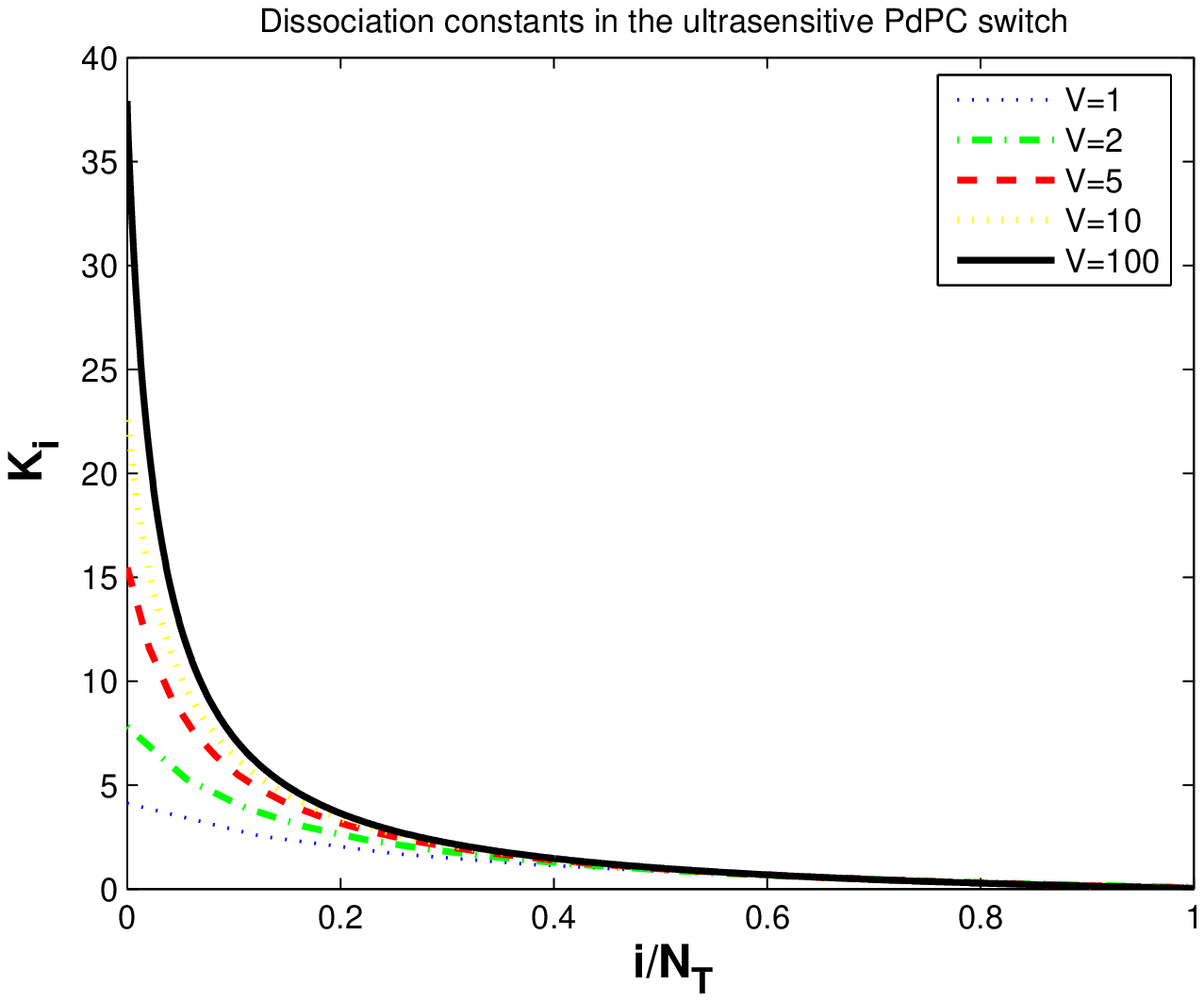}}
\caption[Fig_ultra_switch_coop]{} \label{Fig_ultra_switch_coop}
\end{figure}

\begin{figure}[h]
\centerline{\includegraphics[width=2in,height=4in,angle=270]{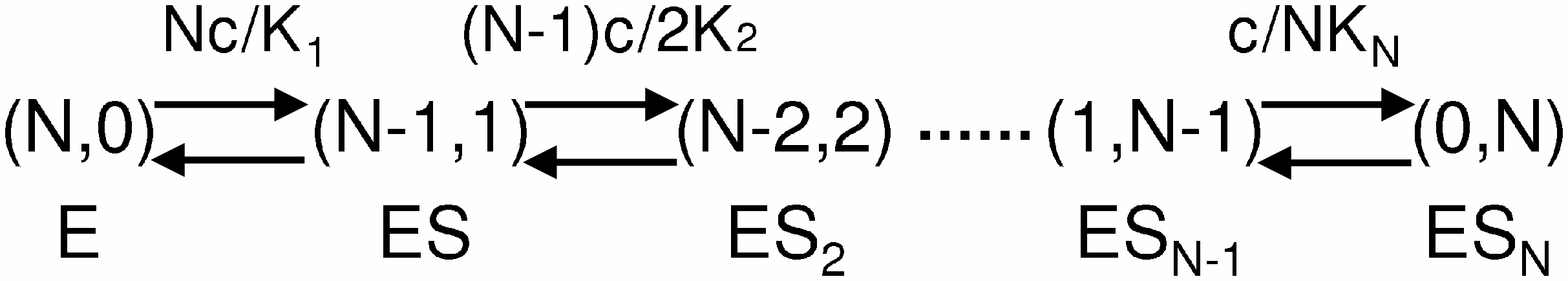}}
\caption[fig_allos_model]{} \label{fig_allos_model}
\end{figure}

\begin{figure}[h]
\centerline{\includegraphics[width=2.5in,height=3in]{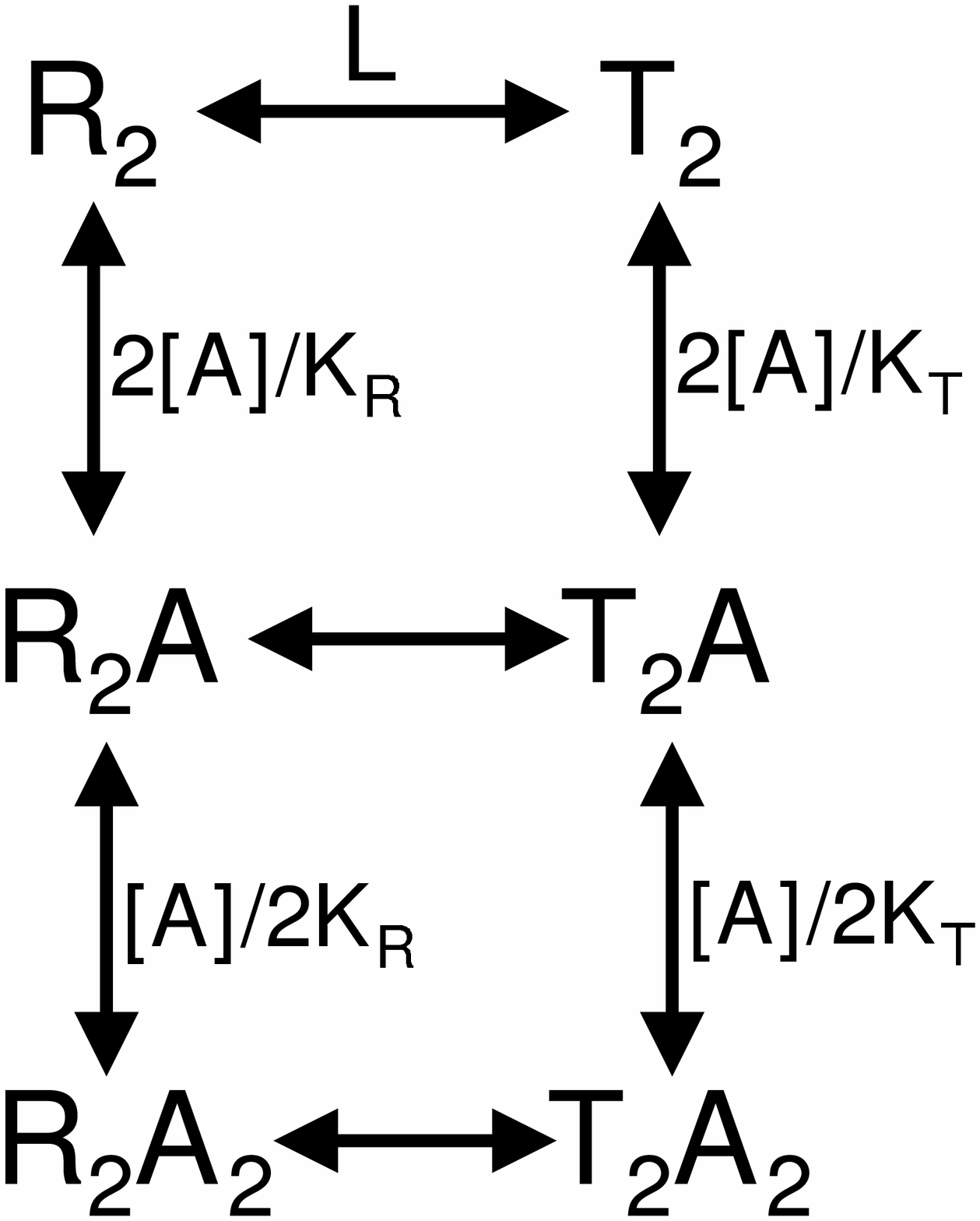}}
\caption[fig5]{} \label{fig5}
\end{figure}

\begin{figure}[h]
\centerline{\includegraphics[width=2.5in,height=3in]{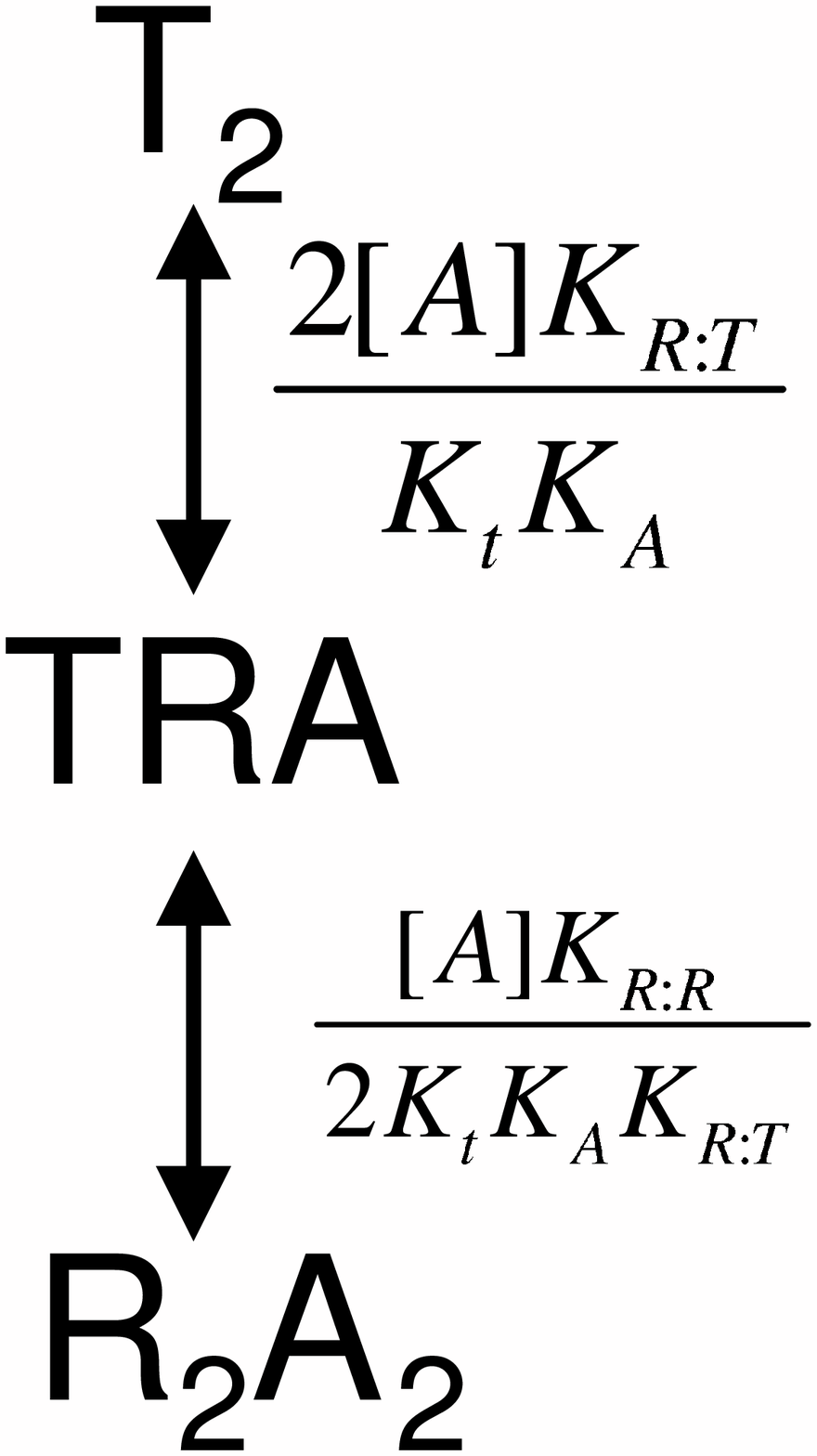}}
\caption[fig7]{} \label{fig7}
\end{figure}

Fig 6: {\sl The curve of $\langle \phi \rangle $ with respect to
$E_{1T}$ at different volumes in the stochastic model of
ultrasensitive PdPC switch under the zero-order approximation, where
the other parameters are the same as that in Fig.
\ref{Fig_ultra_switch_coop}.}

Fig 7: {\sl The curve of $\phi $ with respect to $E_{1T}$ in the
deterministic model of the ultrasensitive PdPC switch without the
zero-order approximation, where the other parameters are the same as
that in Fig. \ref{Fig_ultra_switch_coop}.}

Fig 8: {\sl The curve of $\langle \phi \rangle $ with respect to
$E_{1T}$ of different volumes in the stochastic model of the
ultrasensitive PdPC switch without the zero-order approximation,
where the other parameters are the same as that in Fig.
\ref{Fig_ultra_switch_coop}.}

Fig 9: {\sl The dissociation constants in the ultrasensitive PdPC
switch, where $a_1=10; d_1=1; k_1=1.5; q_1=0.0001; E_{1T}=0.01;
a_2=10; d_2=1; k_2=1.5; q_2=0.0001; E_{2T}=0.01; W_T=10$; and
$\alpha=V_1/V_2^*$. The volume $V$ takes values as 1, 2, 5, 10 and
100, and the molecule number $N=W_TV$ are 10, 20, 50, 100 and 1000
respectively.}

Fig 10: {\sl General model of the allosteric cooperative phenomenon,
where $E$ is the enzyme, $S$ is the substrate and $c=[S]$.}

Fig 11: {\sl Symmetric model for a two-site protein.}

Fig 12: {\sl Sequential model of a two-site protein.}

Fig 13: {\sl Sequential model of quaternary structure.}

\renewcommand{\baselinestretch}{1.4}

\newpage
\newpage

\newpage
\newpage

\newpage

\newpage

\newpage
\newpage

\newpage
\newpage

\newpage
\newpage
\newpage
\begin{figure}[h]
\centerline{\includegraphics[width=2in,height=4in,angle=270]{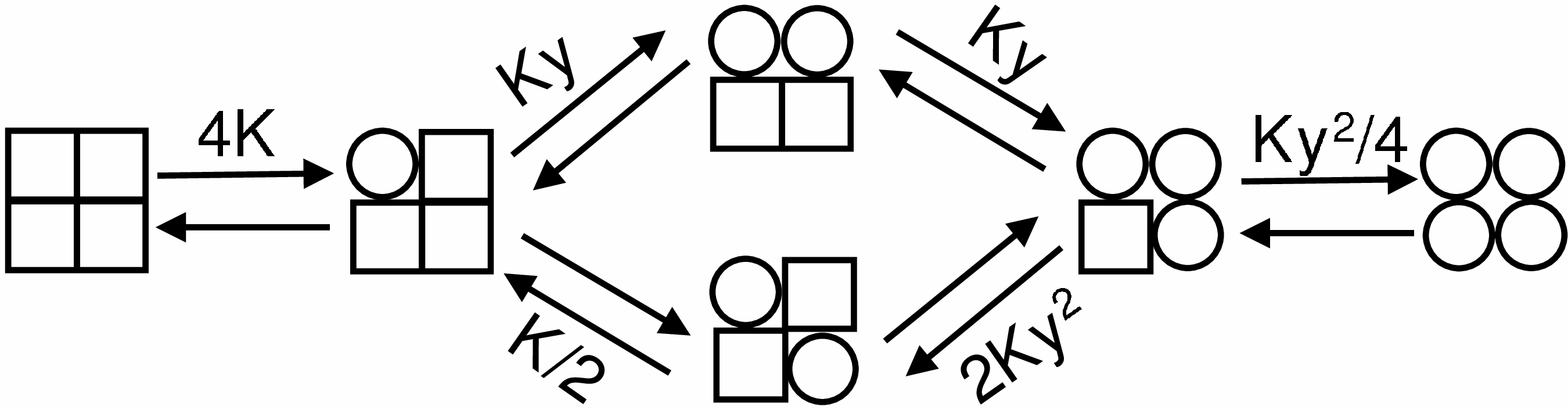}}
\caption[Fig_KNF]{} \label{Fig_KNF}
\end{figure}
\newpage
\begin{table}[h]
    \begin{center}
    \begin{tabular}{|c|c|c|c|}
    \hline
    &\multicolumn{2}{|c|}{Standard Models}&Dissociation constants\\
    \hline
    Temporal&\multicolumn{2}{|c|}{General model(Fig.  \ref{fig_Reduced_stoch})}&$K_i=\frac{(N+1-i)f_2(i/V)}{if_1((N+1-i)/V)}$\\
    \cline{2-4}
    cooperativity&\multicolumn{2}{|c|}{Simple PdPC switch model}&$K_i\approx\frac{\frac{V_2^*}{K_2^*}+\frac{V_1^*}{K_1^*}}{\frac{V_2}{K_2}+\frac{V_1}{K_1}}$\\
    \cline{2-4}
    &\multicolumn{2}{|c|}{Ultrasensitive PdPC switch model}&$K_i\approx \frac{N+1-i}{i}\frac{V_1^*+V_2^*}{V_1+V_2}$\\
    \hline
    Allosteric&Symmetric&Two sites&$K_1=\frac{1+L}{\frac{1}{K_R}+\frac{L}{K_T}},~K_2=\frac{\frac{1}{K_R}+\frac{L}{K_T}}{\frac{1}{K_R^2}+\frac{L}{K_T^2}}$\\
    \cline{3-4}
    cooperativity&model&$N$ sites&$K_i=\frac{\frac{1}{K_R^{i-1}}+\frac{L}{K_T^{i-1}}}{\frac{1}{K_R^{i}}+\frac{L}{K_T^{i}}}$\\
    \cline{2-4}
    &Sequential&Dimer&$K_1=\bar{K}/c,~K_2=c\bar{K}$\\
    \cline{3-4}
    &model&Quaternary&$K_i=\frac{1}{K},~\frac{3}{(2y+1)K},~\frac{2y+1}{3y^2K},~\frac{1}{Ky^2}$\\
    \cline{2-4}
    \hline
    \end{tabular}
    \end{center}
    \caption[tab1]{Summary: a compare of temporal and allosteric cooperativity models through dissociation constants.} \label{tab1}
    \end{table}

\small
\begin{thebibliography}{99}

\bibitem{Fisch71} E.H. Fischer, L.M.G. Heilmeyer,  and R.H. Haschke, {Curr. Top. Cell. Regul.} {\bf 4}, 211 (1971)

\bibitem{Kreb71} E.G. Krebs, {Curr. Top. Cell. Regul.} {\bf 18}, 401 (1980)

\bibitem{HillAV} A.V. Hill, {Journal of Physiology} {\bf 40}, iv (1910)

\bibitem{MWC} J. Monod, J. Wyman, and J.P. Changeux, {J. Mol. Biol.} {\bf 12}, 88 (1965)

\bibitem{KNF} D.E. Koshland Jr., G. Nemethy, and D. Filmer, {Biochmistry} {\bf 5}, 365(1966)

\bibitem{GK} A. Goldbeter, and D.E. Koshland Jr.,
{Proc. Natl. Acad. Sci. USA} {\bf 78}, 6840 (1981)

\bibitem{KGS} D.E. Koshland Jr.,  A. Goldbeter, and J.B. Stock,
{Science} {\bf 217}, 220 (1982)

\bibitem{HF} C.F. Huang,  and J.E. Ferrell Jr.,
 {Proc. Natl. Acad. Sci. USA} {\bf 93}, 10078 (1996)

\bibitem{QH2003} H. Qian,
{Biophys. Chem.} {\bf 105}, 585 (2003)

\bibitem{QH2007} H. Qian,
 {  Annu. Rev. Phys. Chem.} {\bf 58}, 113
(2007)

\bibitem{NP} G. Nicolis,  and  I. Prigogine, {\em Self-organization in nonequilibrium
   systems: from dissipative structures to order through fluctuations.}
   (New York: Wiley 1977)

\bibitem{SC} E.R. Stadtman,  and P.B. Chock,
 {Proc. Natl. Acad. Sci. USA}
{\bf 74}, 2761 (1977)

\bibitem{HNR} R. Heinrich, B.G. Neel,  and T.A. Rapoport:
{Mol. Cell} {\bf 9}, 957 (2002)

\bibitem{Mu} J.D. Murray, {\em Mathematical biology, 3rd Ed.}
(New York: Springer 2002)

\bibitem{FMWT} C.P. Fall, E.S. Marland, J.M. Wagner,  and J.J. Tyson, {\em Computational cell biology.}
 (New York: Springer-Verlag  2002)

\bibitem{Gi76} D.T. Gillespie, {J. Comp. Phys.} {\bf 22}, 403 (1976)

\bibitem{Mc1963} D.A. McQuarrie, {J. Chem. Phys.} {\bf 38}, 437 (1963)

\bibitem{Mc1964}  C.J. Jachimowski, D.A. McQuarrie,  and M.E. Russell, {Biochemistry} {\bf 3}, 1732 (1964)

\bibitem{Mc} D.A. McQuarrie, {J. Appl. Prob.} {\bf 4}, 413 (1967)

\bibitem{GHO} H. Grabert, P. Hanggi,  and  I. Oppenheim, Physica  {\bf l17A}, 300 (1983)

\bibitem{Van} N.G. Van Kampen,  {\em Stochastic Processes in Physics and Chemistry.}
 (Amsterdam: North-Holland 1981)

\bibitem{QH1} H. Qian, S. Saffarian  and E.L. Elson,  {Proc. Natl. Acad. Sci. USA} {\bf 99}, 10376 (2002)

\bibitem{Zhou} T.S. Zhou, L.N. Chen, and R.Q. Wang,
 {Physica D} {\bf 211}, 107 (2005)

\bibitem{SES} P.S. Swain, M.B. Elowitz,  and E.D. Siggia,
{Proc. Natl. Acad. Sci. USA} {\bf 99}, 12795 (2002)

\bibitem{ZMTW} J.S. van Zon, M.J. Morelli, S. Tanase-Nicola,  and
 P.R. ten Wolde, {Biophys. J.} {\bf 91}, 4350 2006

\bibitem{hqjpcm05}
H. Qian, {J. Phys. Cond. Matt.} {\bf 17}, S3783 (2005)

\bibitem{hqjpc06}
H. Qian, {J. Phys. Chem. B.} {\bf 110}, 15063 (2006)

\bibitem{GQQ_MBS2007} H. Ge, H. Qian,  and M. Qian, {
Math. Biosci.} {\bf 211}, 132 (2007)

\bibitem{Ge_JPC2007} H. Ge, {J. Phys. Chem. B} {\bf 112}, 61 (2007)

\bibitem{BQ2007} D.A. Beard,  and H. Qian, {\em Chemical Biophysics: Quantitative Analysis of Cellular
Systems.}
{Cambridge Texts in Biomedical Engineering} (Cambridge University
Press 2008)

\bibitem{Ho} J. Howard, {\em Mechanics of motor proteins and the cytoskeleton.} (Sunderland, MA: Sinauer
2001)

\bibitem{CB} A. Cornish-Bowden, {\em Fundamentals of enzyme kinetics.} 3nd ed.
   (London: Portland Press 2004)

\bibitem{Elf2003} J. Elf, J. Paulsson, O.G. Berg,  and M. Ehrenberg,
{Biophys. J.} {\bf 84}, 154 (2003)

\bibitem{Kur} T.G. Kurtz,
   {  J. Chem. Phys.} {\bf 57}, 2976 (1972)

\bibitem{BPE} O.G. Berg, J. Paulsson,  and M. Ehrenberg,
 {  Biophys. J.} {\bf 79}, 1228 (2000)

\bibitem{MCJ} J. Monod,  J.P. Changeux, and  F. Jacob, {  J. Mol. Biol.} {\bf 6}, 306 (1963)

\bibitem{Ad} G.S. Adair, {  J. Biol. Chem.} {\bf 63}, 529 (1925)

\bibitem{QH_BJREV2008} H. Qian,  and J.A. Cooper, {Biophys. J.} {\bf 47}, 2211 (2008)

\bibitem{RC1987} J. Ricard,  and  A. Cornish-Bowden,
 {Eur. J. Biochem.}
{\bf 166}, 255 (1987)

\bibitem{Kosh58} D.E. Koshland Jr.,
 {  Proc. Natl. Acad. Sci. USA} {\bf 44}, 98 (1958)

\bibitem{Kosh59a} D.E. Koshland Jr., ``Mechanisms of transfer enzymes.'' 305-306 in {\em The Enzymes.} 2nd ed.(Edited by Boyer, P.D.,
Lardy, H. and Myrback, K.) volume 1, Academic Press, New York 1959

\bibitem{Kosh59b} D.E. Koshland Jr.,  {J. Cell. Comp. Physiol.}
 {\bf 54}, supplement 1, 245 (1959)

\bibitem{Pau35} L. Pauling, {\em  Proc. Natl. Acad. Sci.} {\bf 21}, 186 (1935)

\bibitem{Kosh98} D.E. Koshland Jr.,  {Science}
 {\bf 280}, 852 (1998)

\bibitem{HHLM} L.H. Hartwell, J.J. Hopfield, S. Leibler,  and A.W. Murray,  {Nature(London)}
{\bf 402}, C47 (1999)

\bibitem{Fisch55} E.H. Fischer,  and E.G. Krebs, {J. Mol. Chem.} {\bf 216}, 121 (1955)

\bibitem{WD} D.J. Wilkinson,  {\em Stochastic Modelling for Systems
Biology.}
 (Chapman and Hall/CRC 2006) p. 147

\bibitem{JQQb}  D.Q. Jiang,  M. Qian, and M.P. Qian,  {\em Mathematical theory of
   nonequilibrium steady states - On the frontier of probability and dynamical
   systems.} Lect. Notes Math. {\bf 1833} (Berlin:  Springer-Verlag
   2004) Chap.2

\bibitem{Sc} J. Schnakenberg, {Rev. Modern Phys.} {\bf 48},
   571 (1976)
\end{thebibliography}
\end{document}